\documentclass[aps,prx,twocolumn,showpacs,superscriptaddress,amsmath,amssymb]{revtex4-1}

\usepackage{graphicx}
\usepackage[colorlinks=true,linkcolor=blue,citecolor=blue,urlcolor=blue]{hyperref}
\usepackage{bbold}
\usepackage{gensymb}

\def \a {\alpha}
\def \b {\beta}
\def \d {\delta}
\def \D {\Delta}

\def \g {\gamma}

\def \l {\lambda}
\def \L {\Lambda}
\def \o {\omega}

\def \t {\theta}

\def \s {\sigma}

\def \dag {\dagger}

\def \p {\partial}
\def \dge {\degree}

\def \apx {\approx}

\def \til {\tilde}

\def \dag {\dagger}

\newcommand{\intv}[1]{\int_{\mbf #1}}

\def \rar {\rightarrow}

\def \la {\langle}
\def \ra {\rangle}
\def \fr {\frac}
\def \lf {\left}
\def \ri {\right}

\newcommand{\epvl}[1]{\la#1\ra}

\def \Tr {\mathrm{Tr}}
\def \bece {\begin{center}}
\def \ence {\end{center}}
\def \beeq {\begin{equation}}
\def \eneq {\end{equation}}
\def \beal {\begin{aligned}}
\def \enal {\end{aligned}}
\def \bega {\begin{gathered}}
\def \enga {\end{gathered}}
\def \benu {\begin{enumerate}}
\def \ennu {\end{enumerate}}
\def \beit {\begin{itemize}}
\def \enit {\end{itemize}}
\def \bede {\begin{description}}
\def \ende {\end{description}}
\def \betb {\begin{tabular}}
\def \entb {\end{tabular}}
\def \bear {\begin{array}}
\def \enar {\end{array}}

\def \mbf {\mathbf}
\def \mbb {\mathbb}

\def \mca {\mathcal}

\def \bsb{\boldsymbol}
\def \txt {\text}

\begin{document}


\title{A chiral twist on the high-$T_c$ phase diagram in Moir\'e heterostructures}

\author{Yu-Ping Lin}
\affiliation{Department of Physics, University of Colorado, Boulder, Colorado 80309, USA}
\author{Rahul M. Nandkishore}
\affiliation{Department of Physics, University of Colorado, Boulder, Colorado 80309, USA}
\affiliation{Center for Theory of Quantum Matter, University of Colorado, Boulder, Colorado 80309, USA}

\date{\today}

\begin{abstract}
We show that the large orbital degeneracy inherent in Moir\'e heterostructures naturally gives rise to a `high-$T_c$' like phase diagram with a chiral twist - wherein an exotic {\it quantum anomalous Hall} insulator phase is flanked by chiral $d+id$ superconducting domes. Specifically, we analyze repulsively interacting fermions on hexagonal (triangular or honeycomb) lattices near Van Hove filling, with an $\txt{SU}(N_f)$ flavor degeneracy. This model is inspired by recent experiments on graphene Moir\'e heterostructures. At this point, a nested Fermi surface and divergent density of states give rise to strong ($\ln^2$) instabilities to correlated phases, the competition between which can be controllably addressed through a combination of weak coupling parquet renormalization group and Landau-Ginzburg analysis. For $N_f=2$ (i.e. spin degeneracy only) it is known that chiral $d+id$ superconductivity is the unambiguously leading weak coupling instability. Here we show that $N_f\geq4$ leads to a richer (but still unambiguous and fully controllable) behavior, wherein at weak coupling the leading instability is to a fully gapped and chiral {\it Chern insulator}, characterized by a spontaneous breaking of time reversal symmetry and a quantized Hall response. Upon doping this phase gives way to a chiral $d+id$ superconductor. We further consider deforming this minimal model by introducing an orbital splitting of the Van Hove singularities, and discuss the resulting RG flow and phase diagram. Our analysis thus bridges the minimal model and the practical Moir\'e band structures, thereby providing a transparent picture of how the correlated phases arise under various circumstances. Meanwhile, a similar analysis on the square lattice predicts a phase diagram where (for $N_f>2$) a nodal staggered flux phase with `loop current' order gives way upon doping to a nodal $d$-wave superconductor.
\end{abstract}

\maketitle

\section{Introduction}

{\it Chiral} phases of quantum matter spontaneously break time reversal symmetry and exhibit a wealth of fascinating properties, including quantized Hall effects and optical activity, that make them uniquely interesting for both fundamental and technological reasons \cite{Volovik, Haldane, Sigrist}. While {\it insulating} chiral phases are believed to have been found in magnetic topological insulators \cite{Chang}, and {\it superconducting} chiral phases may have been observed in various strontium based materials \cite{Mackenzie, SrPtAs}, the search is still on for a system which can be controllably tuned between insulating and superconducting chiral phases. Meanwhile on the theory level, the search is still on for general principles regarding how to stabilize chiral phases of matter, particularly in systems of correlated electrons. In this work, we show that Moir\'e heterostructures, a system of choice for modern nanoscience, should provide a material platform that can be {\it controllably} tuned between chiral insulating and superconducting phases. Our work also provides insights into how to stabilize chiral phases in systems of correlated electrons. 

\begin{figure}[t]
\centering
\includegraphics[scale = 1]{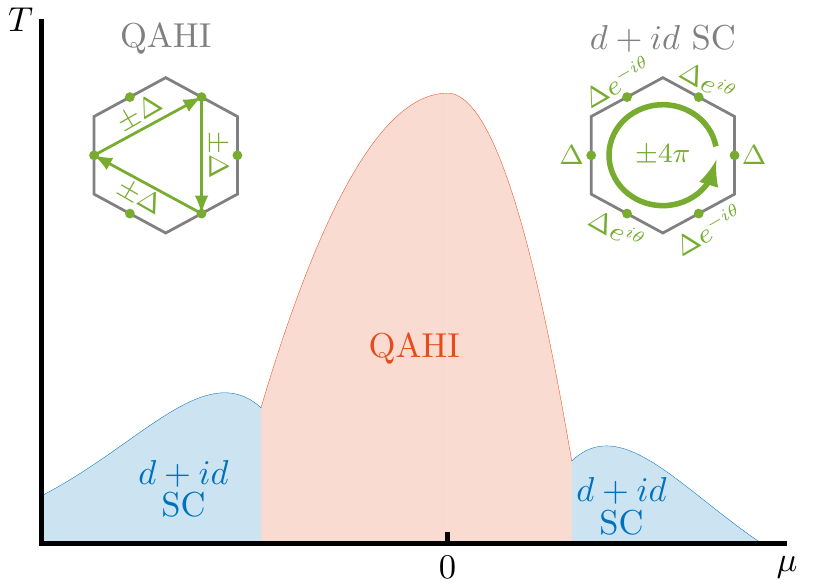}
\caption{\label{fig:phase} Phase diagram for repulsively interacting fermions on hexagonal lattices near Van Hove filling with $\txt{SU}(N_f)$ flavor symmetry, $N_f\geq4$. A quantum anomalous Hall insulator dominates near Van Hove filling $\mu=0$. The state arises from a chiral $3Q$ loop current order with ordering at all nesting momenta. Upon doping this gives way to a chiral $d+id$ superconductor. The phase of the order parameter winds by $\pm4\pi$ around the Fermi surface, where $\t=\pm2\pi/3$ is defined.}
\end{figure}

The study of correlated electrons has been a central theme of condensed matter research for decades. A central open problem in this field is understanding the phase diagram of the cuprate high-$T_c$ superconductors \cite{LeeRMP}, in which a (non-chiral) insulating phase is flanked by domes of (nodal) $d$-wave superconductor. The whole phase diagram is widely believed to originate from a microscopic model of repulsively interacting fermions \cite{LeeRMP}. Recently a new direction has been opened in this field by experiments on graphene Moir\'e heterostructures, such as twisted bilayer graphene \cite{tbg1, tbg2, yankowitz19sci, kerelsky18ax}, or ABC-stacked trilayer graphene on hexagonal boron nitride \cite{3lg}. In these systems there arises a superlattice potential, such that the low energy physics in the reduced Brillouin zone is described by a system of relatively flat bands with Berry curvature \cite{Mele, Bistritzer, Barticevic, KumarNandkishore, SongLevitov, Koshino}, and with numerous fermion flavors from the multiorbital nature \cite{model1,po18prx, model2, model3, model4, zhang19prb,zhang1809ax}. The Fermi level appears to be close to a Van Hove singularity \cite{model3}. Interactions then generate various correlated phases, in a system with unprecedented experimental control. Furthermore, the experimentally observed phase diagram \cite{tbg2} is reminiscent of `high-$T_c$', with an insulating phase flanked by superconducting domes, and with a relatively high ratio of critical temperature to Fermi energy. Of course, there are also important differences to the cuprates, such as the lattice having hexagonal rather than square symmetry, the presence of a large flavor number, and the weak insulating gaps near $0.3~\txt{meV}$ compared to the bandwidths and interactions at the order of $10~\txt{meV}$ \cite{tbg1}.

Motivated by the experimental observations, we consider a system of repulsively interacting fermions on a hexagonal lattice, with the Fermi level close to the Van Hove singularity, and with an $\txt{SU}(N_f)$ flavor degeneracy. Experimental systems are typically in the {\it moderate coupling} regime, which poses challenges for theoretical study. Here we adopt the {\it weak coupling} approach, in the hope that this may yield insight into moderate coupling phenomena. Fermiology is important in the weak coupling framework. In particular, the patches in the immediate vicinity of Van Hove singularities dominate the weak coupling instabilities. We assume the Van Hove singularities to occur at the zone boundary, as appropriate for hexagonal lattices doped to the $M$ point. The Fermi surface is highly nested, giving rise to weak coupling instabilities in multiple channels, the energy scales for which are enhanced by the Van Hove singularities.

The competition between various ordering tendencies is treated in an unbiased manner through a parquet renormalization group (RG) procedure. Whereas for $N_f=2$ (spin degeneracy only) the leading instability at Van Hove filling is known to be in a doubly degenerate $d$-wave superconducting channel \cite{nandkishore12np} (with Landau-Ginzburg analysis resolving the degeneracy in favor of chiral $d+id$ order), for $N_f\geq4$ we show that the leading weak coupling instability shifts, to a triply degenerate channel with imaginary charge density wave (loop current) order. Landau-Ginzburg analysis reveals that the ground state has `triple-$Q$' order, and corresponds to a fully gapped `quantum anomalous Hall' (QAH) phase \cite{Haldane,venderbos16prb} (also known as a Chern insulator), which is a chiral insulator that spontaneously breaks time reversal symmetry. Upon doping, the leading instability shifts to a chiral $d+id$ superconductor. We therefore obtain a `high-$T_c$' like phase diagram as presented in Fig.~\ref{fig:phase}, but with a chiral twist - viz. both the superconducting and the insulating phases are chiral, and spontaneously break time reversal symmetry. These phases are triggered by the flavor fluctuations, which are inherent at low energy \cite{schulz}. We further introduce the orbital splitting to the minimal model, which usually happens in the practical graphene Moir\'e heterostructures. We discuss the resulting RG flow and phase diagram, hence bridging the gap between the minimal model and the practical Moir\'e band structures. We also discuss the behavior of square lattice systems near Van Hove filling with $\txt{SU}(N_f)$ flavor degeneracy. At $N_f>2$, a non-chiral nodal staggered flux phase with loop current order \cite{affleck88prb,nayak00prb,chakravarty01prb,honerkamp04prl} is flanked by a non-chiral $d$-wave nodal superconductor \cite{schulz, dzyaloshinski, furukawa}.

Our work develops naturally from the parquet studies of correlated electrons. Such studies were first developed for nearest neighbor hopping models on square lattices \cite{schulz, dzyaloshinski, furukawa} wherein at half filling there are numerous weak coupling instabilities, with at leading order a degeneracy between (nodal) superconducting and antiferromagnetic orders. A consideration of subleading corrections resolves the degeneracy in favor of antiferromagnetic order, with the antiferromagnet giving way upon doping to superconductivity. The analyses were generalized to honeycomb and triangular lattices in Ref.~\onlinecite{nandkishore12np} where it was shown that at the $M$ point, the leading instability was unambiguously in a doubly degenerate $d$-wave superconducting channel, giving rise to $d+id$ order (a prediction potentially confirmed by experiments on $\txt{SrPtAs}$ \cite{SrPtAs}). In all these works, only spin degeneracy was considered. Our work extends these analyses in a new direction by introducing an enlarged {\it flavor} degeneracy, and finds rich results accordingly. Significantly, the leading instability arises in an imaginary charge density wave channel. While the imaginary density waves are usually subleading and missed in most parquet RG studies (except in e.g. Ref.~\onlinecite{chubukov08prb,chubukov16prx}), our work captures these channels and finds remarkable results accordingly.

Our minimal model analysis adopts the `minimal' setup of Van Hove fermiology, with a particular focus on the patches in the immediate vicinity of Van Hove singularities. This setup is exact in the $\txt{SU}(N_f)$ symmetric minimal models. While the $N_f=4$ models have been adopted in previous phenomenological works on twisted bilayer graphene \cite{xu18prl,model5,kennes18prb,LinKL}, an exact fitness has also been proposed for the topmost valence band of twisted bilayer hexagonal boron nitride \cite{xian18ax}. Note that the breakdown of orbital degeneracy has been pointed out for current graphene Moir\'e heterostructures, including twisted bilayer graphene and ABC-stacked trilayer graphene on hexagonal boron nitride \cite{model1,po18prx, model2, model3, model4,zhang19prb,zhang1809ax}. Nevertheless, our analysis further adopts the splitting between Van Hove singularities and indicates clearly how the pop-up of correlated phases can occur. The general setup of patch fermiology suggests the broadness of potential applicability to the Moir\'e heterostructures. Meanwhile, high flavor degeneracy has also been achieved with ultracold atoms, especially the alkaline-earth family \cite{cazalilla14rpp}. The $\txt{SU}(N_f)$ flavor degeneracy with $N_f>2$ suggests these systems as potential platforms where our minimal model results apply directly. Future studies of, for example, currently unknown Moir\'e heterostructures or other multiorbital materials, may also find useful hints from our work.

Various approaches have been utilized to resolve the moderate coupling problem in Moir\'e heterostructures. Our work differs from most other investigations (e.g. Refs.~\onlinecite{po18prx,xu18prl, model5}), in that most authors have started from {\it strong coupling}. We in contrast have started from weak coupling, which allows us to treat the interaction driven instabilities in an unbiased and controlled manner, and have used the Van Hove singularity to enhance the temperature scales of the instabilities. This approach is justified by the weak insulating gaps observed in experiments, and is also suggested by the increasing bandwidth away from the magic angles. Of course, there is no guarantee that a weak coupling analysis is necessarily appropriate to describe any particular experimental system, but the cleanness and tractability of the analysis, and the remarkable results, make the model interesting in its own right, and may provide a good guide to the behavior of some Moir\'e heterostructures.

Of previous works that have taken a weak coupling approach, our analysis differs from Refs.~\onlinecite{gonzalez19prl, LinKL, xie18ax} in that it works close to Van Hove filling, and takes full account of the competition between various orders. It also differs from the random phase approximation (RPA) analyses \cite{PRL,you18ax}, in that RG treats the intertwinement between various orders unbiasedly while RPA does not. Of the previous RG works, our work differs from the numerically implemented functional RG investigation of Ref.~\onlinecite{kennes18prb}, in that our analytic treatment isolates the most divergent diagrams in a manner that (unlike Ref.~\onlinecite{kennes18prb}) is asymptotically exact at weak coupling. Meanwhile, our minimal model analysis differs from the previous parquet RG analyses Refs.~\onlinecite{isobe18prx,sherkunov18prb} in several important ways. Firstly, unlike Refs.~\onlinecite{isobe18prx,sherkunov18prb} we consider the case where the Van Hove singularities occur at the $M$ points on the zone boundary (as is natural for tight binding models on hexagonal lattices, when far neighbor hoppings are neglected). When Van Hove singularities occur at the $M$ points, then {\it Umklapp} scattering neglected in Refs.~\onlinecite{isobe18prx,sherkunov18prb} have to be taken into account, leading to a very different structure of the RG equations. Secondly, we assume a full $\txt{SU}(N_f)$ symmetry in flavor space, which dramatically simplifies the analysis, and leads to clear, unambiguous results that expose the key features of the problem. Thirdly, our analysis adopts the Fermi surface nesting between each pair of Van Hove singularities. This setup differs from Ref.~\onlinecite{isobe18prx}, where only interorbital nesting is admitted under orbital splitting. Ref.~\onlinecite{sherkunov18prb}, on the other hand, exploits a model where nesting is absent. Lastly, our instability analysis captures the density waves beyond $s$-wave \cite{nayak00prb}, and finds the dominance of loop current order accordingly. When the orbital splitting is introduced, our analysis assumes the $X$-like saddle points and preserves the nesting between each pair of Van Hove singularities. Unlike the $K$-like saddle points in Ref.~\onlinecite{isobe18prx}, our setup captures the fermiology more appropriately in the immediate vecinity of Van Hove singularities \cite{kim16nl}, thereby serves as a more appropriate starting point for an asymptotically exact weak coupling analysis. Note that the phases and phase diagram we obtain are qualitatively different from any of these previous works.

We emphasize also that the Chern insulator predicted from loop current order in our work is qualitatively different from the various QAH phases discussed in graphene Moir\'e heterostructures \cite{PRL,zhang19prb,zhang1809ax,liudai18prb,xie18ax}.  In all of these works, the QAH effect is intimately related to {\it magnetic} order (ordering in the spin-valley space). In contrast, our Chern insulator has no magnetic order, and the QAH effect is of purely orbital origin, much closer in spirit to Haldane's original proposal \cite{Haldane}.

\section{The model and the renormalization group}

\begin{figure}[t]
\centering
\includegraphics[scale = 1]{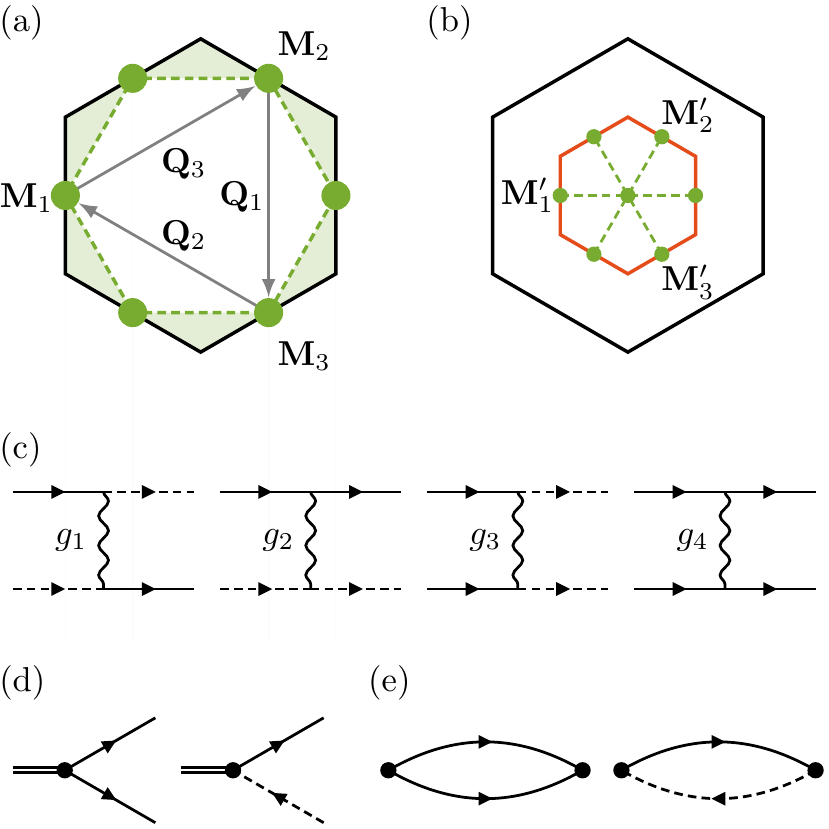}
\caption{\label{fig:bzfd} Setup of patch model. (a) Brillouin zone of hexagonal lattices (black solid) with inscribed Fermi surface (green dashed) at Van Hove filling. The patches are set on the Fermi surface corners $\mbf M_\a$'s. Blanked and shaded regions represent opposite sides of Fermi level. (b) Reduced Brillouin zone in the loop current phase. (c) Four independent interactions in the low energy theory. Solid and dashed lines indicate fermions at different patches. (d) Test vertices in (left) superconducting and (right) density wave channels, with (e) the corresponding  susceptibilities.}
\end{figure}

We consider a system of $N_f$ fermions hopping in a flavor conserving manner on a two dimensional lattice near Van Hove filling. An $\txt{SU}(N_f)$ flavor degeneracy is assumed for the system. The Van Hove singularity (logarithmic divergence in the density of states) arises from $N_p$ saddle points, which are assumed to occur at the Brillouin zone boundary (i.e. at momenta $\bf{M}_{\alpha} = - \bf{M}_{\alpha} $). In triangular and honeycomb lattices $N_p=3$, whereas in square lattices $N_p=2$. However, we develop the analysis for general $N_p$. The low energy theory can be well approximated by a patch model consisting of $N_p$ patches at the saddle points \cite{dzyaloshinski, schulz,furukawa, nandkishore12np}
\beeq
H_0=\sum_{\a=1}^{N_p}\xi_\a\psi_\a^\dag\psi_\a,
\eneq
with the dispersion energies $\xi_\a$'s in the approximate hyperbolic forms (See Appendix~\ref{app:patch}). All $N_p$ inequivalent saddle points are assumed to be mutually nested with nesting momenta $\mbf Q_{\alpha}$'s [see Fig.~\ref{fig:bzfd}(a)]. This situation arises in honeycomb lattices doped to the $M$ point when third neighbor and higher hoppings can be neglected, and also in triangular and square lattices at the appropriate filling when second neighbor and higher hoppings can be neglected. Note that the asymptotically exact weak coupling analysis relies only on the fermiology in the immediate vicinity of Van Hove singularities. Despite the adoption of a minimal model with $\txt{SU}(N_f)$ flavor degeneracy everywhere, the essence lies in the supported fermiology at the Van Hove singularities. Therefore, our analysis applies to any system with the `minimal' setup of Van Hove fermiology, where the Van Hove singularities at $\mbf M_\a$'s carry $\txt{SU}(N_f)$ flavor degeneracy and Fermi surface nesting. The result is independent of the rest details of the system. With the knowledge acquired thereof, we will further show that the systems with orbital splitting can also be interpreted transparently.

The interactions in the low energy theory are assumed to be weakly repulsive, short-ranged, and $\txt{SU}(N_f)$ symmetric. Summarized by Fig.~\ref{fig:bzfd}(c), four inequivalent interactions of similar order exist in the patch model
\beeq
\label{eq:intham}
\beal
H_\txt{int}
&=\fr{1}{2}\sum_{\a\neq\b}\big[g_1\psi_\a^\dag\psi_\b^\dag\psi_\a\psi_\b+g_2\psi_\a^\dag\psi_\b^\dag\psi_\b\psi_\a\\
&\quad+g_3\psi_\a^\dag\psi_\a^\dag\psi_\b\psi_\b\big]+\fr{1}{2}g_4\sum_\a\psi_\a^\dag\psi_\a^\dag\psi_\a\psi_\a,
\enal
\eneq
where the order of fermion flavors is $\s,\s',\s',\s$. Note that the Umklapp scattering $g_3$ is allowed since the nesting momenta satisfy $2\mbf Q=\mbf0$ up to reciprocal lattice vectors.

Our setup parallels the classic works on parquet renormalization group (RG) \cite{dzyaloshinski, schulz, furukawa, nandkishore12np}, except that we have kept the number of patches $N_p$ arbitrary, and have allowed for an $N_f$ flavor degeneracy. As in Refs.~\onlinecite{dzyaloshinski, schulz, furukawa, nandkishore12np}, the divergent density of states and the nested Fermi surface will give rise to divergent susceptibilities in particle-particle and particle-hole channels $\Pi^\txt{pp/ph}_{\mbf q\nu}=\pm T\sum_\o\intv{k}G_{\mbf k\o}G_{(\mp\mbf k+\mbf q)(\mp\o+\nu)}$. Here $G_{\mbf k\o}=(i\o-\xi_{\mbf k})^{-1}$ represents the free fermionic propagator with Matsubara frequency $\o$. We focus on the static part of the susceptibilities and set the bosonic Matsubara frequency $\nu=0$.

Different divergences are manifested in different channels \cite{nandkishore12np}. Due to the Van Hove singularity, two of the channels exhibit $\Pi^\txt{pp}_{\mbf Q},\Pi^\txt{ph}_{\mbf 0}\sim\ln(\L/\max\{T,\mu\})$, where the ultraviolet (UV) cutoff $\L$ is determined by the size of patches, $T$ is the temperature, and $\mu$ is the doping relative to the Van Hove point. The other two susceptibilities receive additional logarithmic divergences. While $\Pi^\txt{pp}_{\mbf 0}$ exhibits the conventional singularity in Cooper channel, $\Pi^\txt{ph}_{\mbf Q}$ acquires a logarithmic divergence from the Fermi surface nesting
\beeq\beal
\Pi^\txt{pp}_{\mbf 0}&\apx h^\txt{pp}\ln\fr{\L}{\max\{T,\mu\}}\ln\fr{\L}{T},\\
\Pi^\txt{ph}_{\mbf Q}&\apx h^\txt{ph}\ln\fr{\L}{\max\{T,\mu\}}\ln\fr{\L}{\max\{T,\mu,t'\}},
\enal\eneq
where $t'$ represents higher neighbor hoppings (third neighbor or higher for honeycomb lattice, second neighbor or higher for square or triangular lattices). The prefactors manifest the characteristic density of states $h^\txt{pp},h^\txt{ph}\sim\nu_0$ and are calculated for square and hexagonal lattices in Appendix~\ref{app:nesting}.

Owing to the logarithmic divergences in susceptibilities, a parquet RG is necessary for the analysis of the low energy theory. The calculations are carried out following Ref.~\onlinecite{nandkishore12np}, making the standard `fast parquet' approximation which focuses on the channels with the most divergent ($\ln^2$) susceptibilities. Starting from the bare UV cutoff $\L$, the shell of fast electron modes is progressively integrated out with decreasing energy or, equivalently, decreasing temperature. Following the spirit of `poor man's scaling', the UV cutoff is not rescaled after each step. The integrated fast modes contribute to the flow of interactions through the $\ln^2$ divergent susceptibilities. This procedure is described by a set of RG equations. We define the dimensionless RG time $y=\Pi^\txt{pp}_{\mbf0}/h^\txt{pp}$, and hence obtain the RG equations for dimensionless couplings $g_i\rar h^\txt{pp}g_i$
\beeq\label{eq:rgeq}\beal
\fr{dg_1}{dy}&=d_1[g_1(2g_2-N_fg_1)+(2-N_f)g_3^2],\\
\fr{dg_2}{dy}&=d_1(g_2^2+g_3^2),\\
\fr{dg_3}{dy}&=2d_1g_3[2g_2-(N_f-1)g_1]-g_3[(N_p-2)g_3+2g_4],\\
\fr{dg_4}{dy}&=-(N_p-1)g_3^2-g_4^2.
\enal\eneq
The nesting parameter $d_1(y)=d\Pi^\txt{ph}_{\mbf Q}/d(h^\txt{pp}y)\apx\Pi^\txt{ph}_{\mbf Q}/\Pi^\txt{pp}_{\mbf 0}$ determines the nesting degree $0\leq d_1(y)\leq d_1^\txt{max}$, where the maximum $d_1^\txt{max}=h^\txt{ph}/h^\txt{pp}$ characterizes the maximal nesting at Van Hove filling. A detailed analysis of maximal nesting on different lattices is demonstrated in Appendix~\ref{app:nesting}. For square lattice $d_1^\txt{max}=1$, indicating that the patches enjoy the perfect nesting of Fermi surface. However, the perfect nesting is not accessible to the patches on hexagonal lattices, and a lower maximum $d_1^\txt{max}=1/2$ is manifested accordingly. Notice that the equations for $g_1$ and $g_3$ depend on $N_f$ due to the involvement of interpatch internal fermion loops. The patch number $N_p$ is present in equations for $g_3$ and $g_4$ since the internal Umklapp scattering contributes. These equations reduce to the square lattice equations of Refs.~\onlinecite{dzyaloshinski, schulz, furukawa} when we set $N_p=2, N_f=2$, and to the hexagonal lattice equations of Ref.~\onlinecite{nandkishore12np} when we set $N_p=3, N_f=2$.

We analyze the RG equations with the setup of bare weak repulsions $g_1,g_2,g_3,g_4\geq0$ and finite nesting $d_1(y)>0$. Motivated by the Moir\'e heterostructures based on graphene and hexagonal boron nitride, the numbers $N_f=4$ and $N_p=3$ are chosen \cite{model1,po18prx, model2, model3, model4, model5, zhang19prb,zhang1809ax,xian18ax,xu18prl,kennes18prb}. Note that under the RG flow, $g_2$ increases monotonically and diverges at a critical scale $y_c$. Meanwhile, $g_3$ remains positive semidefinite, while $g_4$ decreases monotonically and may change sign under the RG flow. The behavior of $g_1$ depends on $N_f$. For $N_f=2$, $g_1$ is positive semidefinite, but for $N_f>2$ it can change sign. A detailed analysis of the RG equations following Ref.~\onlinecite{nandkishore12np} is presented in Appendix~\ref{app:ft}, and reveals that for {\it any} choice of bare repulsive interactions, there is a unique fixed trajectory i.e. as the system flows to strong coupling, the ratios of the couplings tend to specific values.

\begin{figure}[b]
\centering
\includegraphics[scale = 1]{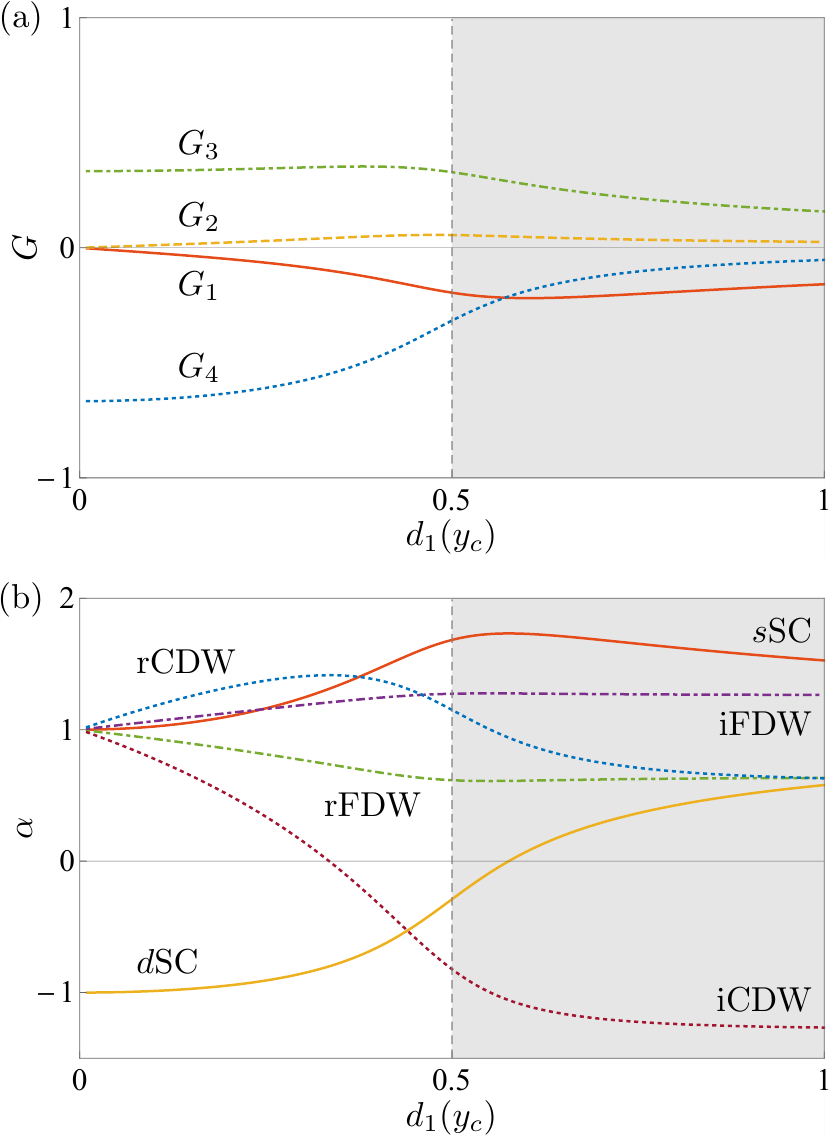}
\caption{\label{fig:rg} Parquet RG results for two orbital hexagonal lattices with $N_f=4$ and $N_p=3$. Notice that only the regime below $d_1^\txt{max}=1/2$ is accessible. (a) Critical interactions $G_i$'s. (b) Susceptibility exponents $\a_I$'s. The most negative exponent indicates the leading instability. While the $d$-wave superconductivity is leading in the low nesting regime, loop current state takes over at high nesting.}
\end{figure}

The fixed trajectory may be determined by making an ansatz of the interactions
\beeq
\label{eq:gans}
g_i(y)=\fr{G_i}{y_c-y}.
\eneq
Substitution into the RG equations yields a set of algebraic equations, which may be straightforwardly solved. Discarding the solutions that cannot be accessed starting from repulsive interactions, and the solutions that are unstable to perturbations, we are left with a unique set of critical interactions $G_i$'s. For single layer graphene $N_f=2$, the fixed trajectories manifest $-G_4>G_3>G_2>G_1=0$ at all nesting $0\leq d_1\leq1$ \cite{nandkishore12np}. However, different features are observed for $N_f=4$ [see Fig.~\ref{fig:rg}(a)]. While $-G_4$ decreases toward zero with increasing nesting, $-G_1$ increases and becomes the largest among all interactions at (inaccessible) perfect nesting $d_1=1$. These features indicate a switch between different fixed trajectories at certain nesting. A transition between different instabilities may also occur accordingly.

\section{Instability analysis}
\label{sec:insana}

To determine the leading instability as the system flows to strong coupling, we introduce the test vertices coupling to the particle-particle and particle-hole bilinears \cite{chubukov08prb,chubukov16prx}
\beeq
\label{eq:instham}
\d H=\sum(\D\psi^\dag\psi^{(\dag)}+\txt{H.c.}).
\eneq
Under the RG flow, these test vertices receive corrections from the particle-particle and particle-hole susceptibilities. Different irreducible channels manifest different flows of test vertices. Based on the test vertex analysis, the instabilities can be inspected through the probe of susceptibilities. The susceptibility that diverges most rapidly under RG represents the leading instability.

\subsection{Test vertex}

We focus on the channels receiving $\ln^2$ divergent susceptibilities - test vertices in channels receiving only $\ln$ divergent susceptibilities do not grow strong before the problem flows to strong coupling \cite{nandkishore12np}. This corresponds to a focus on superconducting and density wave channels [Fig.~\ref{fig:bzfd}(d)]. In the superconducting channels, the test vertices are coupled to intrapatch particle-particle pairings $\D_\a\psi_{\a\s}^\dag\psi_{\a\s'}^\dag$ with $\s>\s'$. These flavor pairings exhibit the antisymmetric $\txt{SU}(N_f)$ irreducible representations. For the density wave channels, interpatch particle-hole pairings $\D_{\a\b}\psi_{\b\s}^\dag\psi_{\a\s'}$ with $\a>\b$ are introduced. Several kinds of irreducible channels can be identified. For the real and imaginary charge density wave (r/iCDW) channels, a summation over all uniform flavor pairings $\sum_\s\psi_{\a\s}^\dag\psi_{\b\s}$ manifests the trivial $\txt{SU}(N_f)$ irreducible representation. For the flavor density wave (r/iFDW) channels, the remaining $\txt{SU}(N_f)$ irreducible representations are relevant.

The corrections to test vertices under RG are described by a set of differential equations. From the analysis in Appendix~\ref{app:inst}, the superconducting (SC) and density wave (DW) channels manifest the equations
\beeq
\label{eq:tvcorrection}
\fr{d\D_\txt{SC}}{dy}=-g_\txt{SC}\D_\txt{SC},\quad
\fr{d\D_\txt{DW}}{dy}=-d_1g_\txt{DW}\D_\txt{DW}.
\eneq
The interactions are linear combinations of the four inequivalent interactions
\beeq
\bega
g_{s\txt{SC}}=(N_p-1)g_3+g_4,\quad g_{d\txt{SC}}=-g_3+g_4,\\
g_{\txt{r/iFDW}}=-(g_2\pm g_3),\\
g_{\txt{r/iCDW}}=N_fg_1-g_2\pm(N_f-1)g_3.
\enga
\eneq
Since the density waves are commensurate with the lattice, the real and imaginary components are decoupled under the RG flow. Along the fixed trajectories, the ansatz of interactions Eq.~(\ref{eq:gans}) indicates the critical scaling of test vertices
\beeq
\label{eq:tvtscaling}
\D_I(y)\sim(y_c-y)^{\b_I}.
\eneq
The exponent $\b_I$ in each channel $I$ is a linear combination of critical interactions
\beeq
\bega
\b_{s\txt{SC}}=(N_p-1)G_3+G_4,\quad \b_{d\txt{SC}}=-G_3+G_4,\\
\b_{\txt{r/iFDW}}=-d_1(G_2\pm G_3),\\
\b_{\txt{r/iCDW}}=d_1[N_fG_1-G_2\pm(N_f-1)G_3].
\enga
\eneq

\subsection{Susceptibility and phase diagram}

A more concrete instability analysis is offered by the probe of susceptibilities in these channels \cite{zanchi00prb,binz02epjb}. Under the RG flow, the perturbing Hamiltonian Eq.~(\ref{eq:instham}) in each channel becomes scale dependent. Up to second order of test vertices $\D$ and $\D^*$, the perturbing Hamiltonian reads
\beeq
\d H_I(y)
=\sum\D_I^\dag(y)\chi_I(y)\D_I(y)+\d H_I^\psi(y),
\eneq
where the $\psi$ dependent terms are collected by $\d H_I^\psi(y)$. Consider the static part of correlation function
\beeq
\chi_I(y)=-T\lf.\fr{\d^2\ln Z(y)}{\d\D_I(0)\d\D_I^*(0)}\ri|_{\D_I^*(0),\D_I(0),\psi^\dag,\psi=0},
\eneq
where $Z(y)$ denotes the partition function at scale $y$. An initial value $\chi_I(0)=0$ is determined by the initial perturbing Hamiltonian Eq.~(\ref{eq:instham}). As a second order derivative of free energy $F=-T\ln Z$ with respect to external fields, the correlation function $\chi_I(y)$ exhibits the proper form of a physical susceptibility. The constraint $\psi^\dag,\psi=0$ indicates that only the response of the outer shells with lower scales $y'<y$ are included. Equivalently, at certain temperature or energy, only the response of electron modes at higher temperatures or energies are measured. Due to this condition, the correlation function $\chi_I(y)$ can be regarded as the physical susceptibility at certain temperature or energy.

Under the RG flow, the susceptibilities receive corrections illustrated by the diagram in Fig.~\ref{fig:bzfd}(e)
\beeq
\label{eq:chicorrection}
\fr{d\chi_\txt{SC}}{dy}=|\D_\txt{SC}|^2,\quad
\fr{d\chi_\txt{DW}}{dy}=d_1|\D_\txt{DW}|^2.
\eneq
With the solution of test vertices Eq.~(\ref{eq:tvtscaling}), the critical scaling of susceptibilities along fixed trajectories is derived
\beeq
\chi_\txt{SC}\sim(y_c-y)^{\a_\txt{SC}},\quad
\chi_\txt{DW}\sim d_1(y_c-y)^{\a_\txt{DW}}.
\eneq
The susceptibility exponents are related to the exponents of test vertices \cite{dzyaloshinski,binz02epjb,cvetkovic12prb,chubukov16prx}
\beeq
\a_I=2\b_I+1.
\eneq
An instability can emerge only when the corresponding susceptibility diverges $\a_I<0$. The leading instability is determined by the most negative susceptibility exponent, since the corresponding susceptibility diverges the most.

\begin{figure}[t]
\centering
\includegraphics[scale = 1]{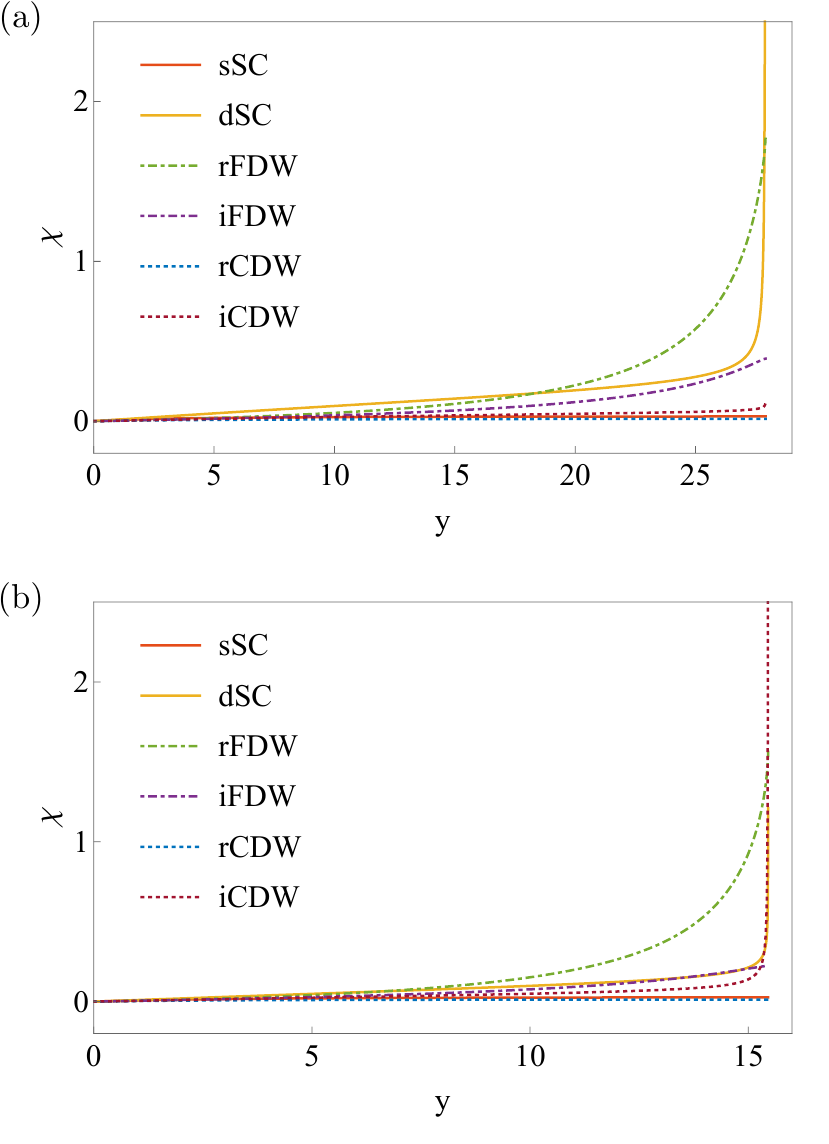}
\caption{\label{fig:chi} Evolution of susceptibilities under RG. The bare interactions $g_i=g_0$ with $g_0=0.1$ are chosen, and $d_1$ is set as constant throughout the RG flow. The leading susceptibility diverges at a scale $y_c\sim g_0^{-1}$, indicating the development of an instability. (a) At $d_1=0.3$, the leading instability occurs in the $d$-wave superconducting channel since only $\chi_{d\txt{SC}}$ diverges. (b) At $d_1=0.5$, both $\chi_\txt{iCDW}$ and $\chi_{d\txt{SC}}$ are divergent. The loop current state is dominant since $\chi_\txt{iCDW}$ grows more rapidly. Notice that at intermediate scales, the real FDW state has the largest susceptibility, but it does not represent the leading weak coupling instability as the problem flows to strong coupling. }
\end{figure}

The susceptibility exponents in the two orbital hexagonal lattice model are presented in Fig.~\ref{fig:rg}(b). The phase diagram exhibits two different phases at different nesting regimes. In the low nesting regime, the $d$-wave superconductivity is dominant as in the single layer graphene. However, as the nesting degree increases, the imaginary CDW state, also known as the loop current state, is enhanced. Above a critical nesting $d_1^c$, the loop current state overcomes the $d$-wave superconductivity and becomes the leading instability. Notice that the phase transition is absent in the single layer graphene, where the $d$-wave superconductivity dominates at all nesting. In the two orbital model, the internal fermion loop is enhanced by extra fermion flavors. With sufficient enhancement at high nesting, the loop current state is triggered, which defeats the $d$-wave superconductivity and causes a phase transition.

The fixed trajectory analysis is verified by probing the evolution of susceptibilities under RG. Setting the bare interactions $g_i=g_0$, the RG flow runs until the critical scale $y_c\sim g_0^{-1}$ is reached. The susceptibilities evolve accordingly as presented in Fig.~\ref{fig:chi}. Among various channels, the real FDW grows first at intermediate scales. This indicates the importance of flavor fluctuations in the low energy theory. Things become different as the RG flow runs further. Through the Umklapp scattering $g_3$, some other channels receive the mediation of flavor fluctuations. The rapid growth of real FDW then leaks out and enhances these channels \cite{schulz}. Approaching the critical scale, the fixed trajectory instabilities surpass the real FDW and diverge. These instabilities manifest enhanced critical temperature and ordering energy
\beeq
T_c,E_0\sim\L\exp(-1/\sqrt{\l}),
\eneq
where the dimensionless coupling $\l\sim g_0\nu_0$ depends on how the bare interactions $g_i$'s and the nesting parameter $d_1(y)$ are modelled. The most divergent susceptibility determines the leading instability. While the $d$-wave superconductivity diverges the most in the low nesting regime, loop current state dominates above the transition nesting $d_1^c$. These results confirm the phase diagram Fig.~\ref{fig:rg}(b) obtained from the fixed trajectory analysis.

Later analysis identifies the loop current phase as a gapped quantum anomalous Hall insulator (QAHI) \cite{Haldane,venderbos16prb}, and the superconductor as a $d+id$ superconductor \cite{nandkishore12np}. The breakdown of time reversal symmetry in both phases indicates a phase diagram composed of two chiral phases, with a phase transition inbetween. Note that the nesting parameter $d_1$ can be controlled by doping away from the saddle point. We therefore expect that close to the Van Hove point, the system will be a (chiral) QAHI, which will give way upon doping to a (chiral) $d+id$ superconductor. This leads to the phase diagram presented in Fig.~\ref{fig:phase}.

We briefly discuss the results for larger flavor number $N_f>4$ \footnote{A QAHI phase is not present for $N_f=3$ since the transition nesting parameter $d_1^c>1/2$ is beyond the available regime.}. In the large flavor regime, the critical interactions $G_i$ reduce as $N_f^{-1}$. This reduction implies the vanishing of most channels at finite nesting, including the $d+id$ superconductivity. However, the QAHI remains robust due to a balancing factor $N_f$ in the susceptibility exponent. Therefore, the transition nesting decreases with increasing flavor number, indicating an expansion of QAHI in the doping phase diagram. This clearly reveals the essential role of flavor degeneracy in stabilizing a (chiral) insulating phase on the hexagonal lattices.

\subsection{How robust are these results?}

We have solved the minimal model problem in the asymptotic weak coupling limit, and obtained answers that are independent of the details of the bare couplings, as long as these are repulsive and sufficiently weak. A question to ask is whether these results are robust against various deviations from our setup and analysis.

\subsubsection{Away from weak coupling}

What if the bare interactions are not so weak? Our RG analysis is asymptotically exact in the limit of weak coupling, such that the problem has time to flow to the fixed trajectory before the strong coupling limit is reached. However, if the couplings are not so weak, then the strong coupling regime may already be entered before the fixed trajectory is reached. In this case the instability realized may differ from the predictions of the asymptotic weak coupling treatment. The results in such a `moderate coupling' regime will depend on the choice of initial `bare' couplings. We note however that starting from `Hubbard' interactions $g_i=g_0$, the real FDW manifests the largest susceptibility at intermediate scales (Fig.~\ref{fig:chi}). Therefore, the real FDW may be a contender if interactions are not so weak. In other words, the flavor fluctuations may dominate the system without triggering the fixed trajectory instabilities through the mediation. Note that the real FDW is reminiscent of the spin density wave (SDW) in single layer graphene \cite{martin08prl,nandkishore12prl,Fernandes,chern12prl,venderbos16prb2}, which manifests the spin fluctuations with $N_f=2$.

The correlated phases at moderate coupling have also been examined by functional RG analyses. With Hubbard repulsions, the trigger of a $d+id$ superconductivity near Van Hove filling is revealed \cite{kennes18prb}. Some weak signals hint the appearance of a density wave beyond $s$-wave, yet the identification of this state is missed. Subsequent to the first posting of our work, another study observes a clear evidence of QAHI by introducing the nearest neighbor $\txt{SU}(4)$ exchange \cite{classen19prb}. Amazingly, the state is robust against and even enhanced by the nearest neighbor exchange. The real FDW, on the other hand, can occur when the spin or orbital Hund's coupling is relevant. However, the state is weak against nearest neighbor exchange as expected, since it is not a fixed trajectory instability. With these functional RG observations, the eligibility of our asymptotically exact RG at moderate coupling is reinforced.

\subsubsection{Away from zone boundary}

What if the Van Hove singularities are shifted away from the zone boundary, e.g. because further neighbor hoppings are not that weak? Any displacement of the Van Hove singularities from the zone boundary introduces an additional energy scale into the RG. If the strong coupling regime is reached before we hit this scale, then our analysis should carry through unchanged. In contrast, if this scale exceeds the original cutoff scale for the RG, then a qualitatively different analysis analogous to Ref.~\onlinecite{isobe18prx} is called for. What if the Van Hove points are displaced by less than the original cutoff scale for the RG (such that our analysis is the correct one at short RG times), but nevertheless by enough that we hit the displacement scale before the problem reaches strong coupling? In this case, the results could change in a manner not captured by either our analysis or Ref.~\onlinecite{isobe18prx}. In particular, while Umklapps will be present above this scale, the $g_3$ channel will be significantly suppressed below this scale due to the prohibition of the Umklapp process $(\mbf M_\a,\mbf M_\a)\rar(\mbf M_\b,\mbf M_\b)$, although there is still the momentum conserving process $(\mbf M_\a,-\mbf M_\a)\rar(\mbf M_\b,-\mbf M_\b)$. A detailed treatment of the case of Van Hove points displaced from the zone boundary is discussed in Sec.~\ref{sec:gmh}.

\subsubsection{Away from fermion bilinears}

We caution that while our RG is asymptotically exact, our instability analysis has considered only the set of phases characterized by order parameters related to condensates of fermion bilinears. These phases manifest local site, bond, or loop current orders in real space. More exotic correlated phases, such as topologically ordered phases with no local order parameter, are beyond the scope of any such approach.

\section{The ordered states are chiral}

We have identified the leading instabilities using parquet RG as an imaginary CDW (loop current) state at Van Hove filling, which gives way upon doping to a $d$-wave superconducting state. However, the imaginary CDW channel is triply degenerate (order can develop along any of the three nesting momenta $\mbf Q_\a$'s), whereas the superconducting channel is doubly degenerate \cite{nandkishore12np}. We now determine the lifting of these degeneracies.

\subsection{Quantum anomalous Hall insulator}

For the loop current channel, there are three available order parameters $\D_\a=\epvl{\psi_\b^\dag\psi_\g}$ at nesting momenta $\mbf Q_\a=\mbf M_\b-\mbf M_\g$ [Fig.~\ref{fig:bzfd}(a)]. These order parameters are purely imaginary $\D_\a=i\txt{Im}\epvl{\psi_\b^\dag\psi_\g}$ and indicates loop current orders in real space. The phase of the order parameters is fixed because the density wave is commensurate with the lattice. An $\mbb R^3$ manifold of imaginary order parameters is thus present. A natural question then arises as what configuration is favored when the ordered phase develops at low temperature. This can be addressed by analyzing the dependence of ordering energy on various order parameters.

In the loop current phase, a breakdown of translational symmetry occurs as in all the other density wave phases. The commensurate momenta $2\mbf Q=0$ implies an enlarged quadrupled unit cell and a reduced Brillouin zone with halved lengths [Fig.~\ref{fig:bzfd}(b)]. Before the loop current orders develop, the noninteracting band structure is obtained by a folding of the original bands. Three nodal lines connecting between opposite edge centers $\pm\mbf M_\a'=\pm\mbf M_\a/2$'s constitute the Fermi surface. A crossing occurs at the zone center $\mbf 0$, leading to a triply degenerate quadratic band crossing point (QBCP). Inheriting the $d$-wave structures of original saddle points $\mbf M_\a$'s, the QBCP carries a nontrivial $2\pi$ Berry flux, and is protected by the combination of $\txt{C}_6$ and time reversal symmetries \cite{sun09prl}.

\begin{figure}[t]
\centering
\includegraphics[scale = 1]{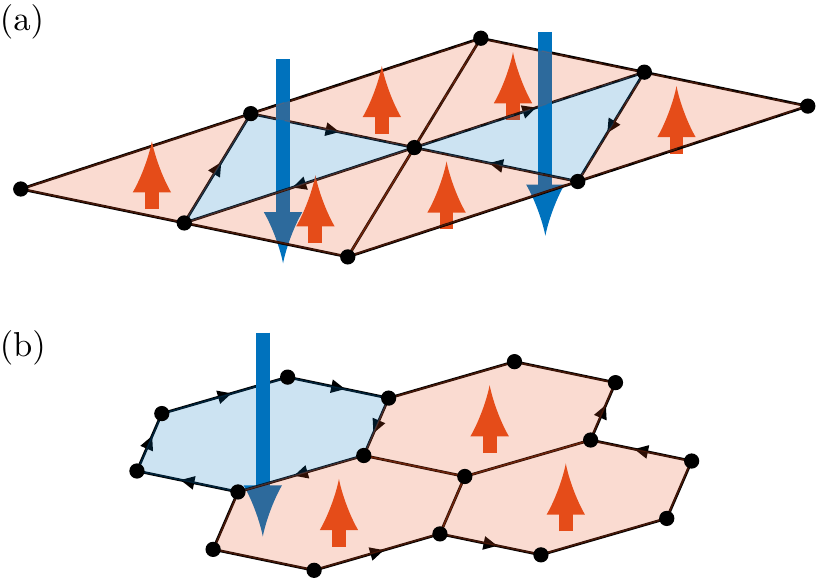}
\caption{\label{fig:qahirs} Real space configurations of quantum anomalous Hall insulator on the (a) triangular and (b) honeycomb lattices, showing pattern of fluxes through the real space quadrupled unit cell. The ratio of $-3\phi$ and $\phi$ fluxes is $1:3$ in each quadrupled unit cell, leading to a zero net flux.}
\end{figure}

When the loop current orders develop, the noninteracting nodal structure is diminished. To analyze the resulting gap structure, the imaginary order parameters are relaxed from the patch model to the whole Brillouin zone $\D_{\a\mbf k}=\epvl{\psi_{\mbf k+\mbf Q_\a}^\dag\psi_{\mbf k}}$. The condition $\D_{\a\mbf M_\g}=\D_{\a(\mbf M_\g+\mbf Q_\a)}^*$ then requires a $d$-wave structure of the order parameter \cite{venderbos16prb}. A simultaneous ordering at all nesting momenta, known as a $3Q$ state, allows to gap the entire Fermi surface, and is expected to maximize the ordering energy. This result establishes the $3Q$ state as the leading order in the loop current phase. A rigorous justification is conducted by a Landau-Ginzburg analysis following Ref.~\onlinecite{nandkishore12prl} in Appendix~\ref{app:lg}. Notice that the $3Q$ states manifest a $\mbb Z_4$ manifold composed of four inequivalent states $(\D,\D,\D)$, $(\D,\D,-\D)$, $(\D,-\D,\D)$, and $(-\D,\D,\D)$. As illustrated in Fig.~\ref{fig:qahirs}, each inequivalent state exhibits a real space flux pattern through the quadrupled unit cell with zero net flux \cite{venderbos16prb}. A $\mbb{Z}_4$ translational symmetry breaking occurs in the $3Q$ manifold when the order develops \cite{nandkishore12prl, Fernandes}.

Note that the $3Q$ loop current state breaks time reversal symmetry and manifests itself as a chiral Chern insulator \cite{chern12prl,venderbos16prb}. We briefly describe how the system is gapped out by the loop current orders, where the Chern insulator emerges as a legacy of QBCP. When an order arises at certain nesting momentum, time reversal symmetry is present up to a translation along the other two nesting momenta. An effective time reversal symmetry thus survives up to the $2Q$ state and keeps the system gapless. As the third order develops, there is no eligible translation, and time reversal symmetry is broken inevitably. The topological charge of original QBCP is transferred onto the resulting gapped bands, thereby triggers a Chern insulator in the $3Q$ state. 
An expected Chern number $C=\pm1$ is confirmed by a Berry flux computation following Ref.~\onlinecite{fukui05jpsp}. The Chern insulator manifests a quantum Hall effect in the absence of external magnetic field, with a quantized Hall conductivity $\s_{xy}=\pm N_fe^2/h$ triggered by the intrinsic fluxes (Fig.~\ref{fig:qahirs}). This phenomenon identifies the chiral $3Q$ loop current state as a quantum anomalous Hall insulator.

\subsection{$d+id$ Superconductivity}

In the $d$-wave superconducting channel, two degenerate patch orders $\D^1=(\D/\sqrt6)(2,-1,-1)$ and $\D^2=(\D/\sqrt2)(0,1,-1)$ are present. Each component describes an intrapatch particle-particle condensate $\D^{1,2}_\a=\epvl{\psi_{\a\s}\psi_{\a\s'}}^{1,2}$ with certain flavor pairing $\s>\s'$. The $d$-wave nature is recognized by relaxing the patch order to the whole Brillouin zone, where the gap functions $\D^{1,2}_{\mbf k}=\epvl{\psi_{-\mbf k\s}\psi_{\mbf k\s'}}^{1,2}$ change signs four times around the Fermi surface. Previous work on single layer graphene \cite{nandkishore12np} has revealed the minimization of free energy at $\D^\pm=\D^1\pm i\D^2$. The state manifests a full gap in the Bogoliubov-de Gennes quasiparticle bands and maximizes the ordering energy. Therefore, the $d+id$ state $\D^\pm=\D(1,\exp[\pm i2\pi/3],\exp[\mp i2\pi/3])$ dominates the $d$-wave superconducting channel. The order parameter $\D^\pm_{\mbf k}$ exhibits a winding around the Fermi surface, where a phase $\pm4\pi$ is acquired after a full winding. Time reversal symmetry is broken accordingly. This corresponds to a chiral superconducting phase exhibiting thermal and spin quantum Hall effects.

Provided the large flavor number $N_f$, the $d+id$ superconductivity has a large degeneracy among the $N_{fp}=N_f(N_f-1)/2$ antisymmetric flavor pairings. The according competition leads to strong fluctuations and suppresses the superconducting order. Mermin-Wagner theorem indicates that the breakdown of continuous $\txt{O}(N_{fp})$ symmetry only occurs at zero temperature \cite{mermin66prl, sachdevbook}. The absence of superconductivity at finite temperature is thus suggested. Nevertheless, the $\txt{O}(N_{fp})$ pairing degeneracy is usually lifted by symmetry breaking perturbations in practical systems. An usually observed example is the spin Hund's coupling in multiorbital systems. Given a spin Hund's coupling in the two orbital model with $N_f=4$ \cite{model1,model3}, previous analysis has uncovered an effective anti-Hund's coupling in the superconducting channels \cite{LinKL}. The sixfold degeneracy of $d+id$ superconductivity is lifted, leaving only one dominant spin singlet channel. Similar effects can also occur in systems with more orbitals. With the symmetry breaking perturbations, a cutoff scale $\D_c$ is introduced to the RG flow in the nonlinear sigma model description \cite{polyakov}. A nonzero critical temperature appears and takes a suppressed form
\beeq
\label{eq:supptc}
T_c\sim\fr{2\pi\rho_s}{(N_{fp}-2)\ln(\D/\D_c)},
\eneq
where $\rho_s$ is the stiffness of flavor pairing fluctuations and $\D$ is the ordering energy scale. Therefore, a $d+id$ superconducting phase can still arise at finite temperature. Note that the critical temperature decreases with increasing flavor number, since the superconducting order is suppressed more severely when more degenerate flavor pairings are present.

\section{Towards graphene Moir\'e heterostructures}
\label{sec:gmh}

We have analyzed the hexagonal lattice minimal model with $\txt{SU}(N_f)$ flavor degeneracy. An asymptotically exact weak coupling analysis uncovers a chiral `high-$T_c$' phase diagram near Van Hove filling. The minimal model setup and the asymptotically exact analysis suggest the potentially broad applicability and extention of our results. Motivated by recent experiments on graphene Moir\'e heterostructures \cite{tbg1, tbg2, yankowitz19sci, kerelsky18ax,3lg}, a crucial issue concerns whether the phase diagram appears in these practical systems. The essential point is how the RG flows and the according instabilities evolve when the Van Hove fermiology deforms from the minimal model.

\subsection{Van Hove fermiology}

\begin{figure}[b]
\centering
\includegraphics[scale = 1]{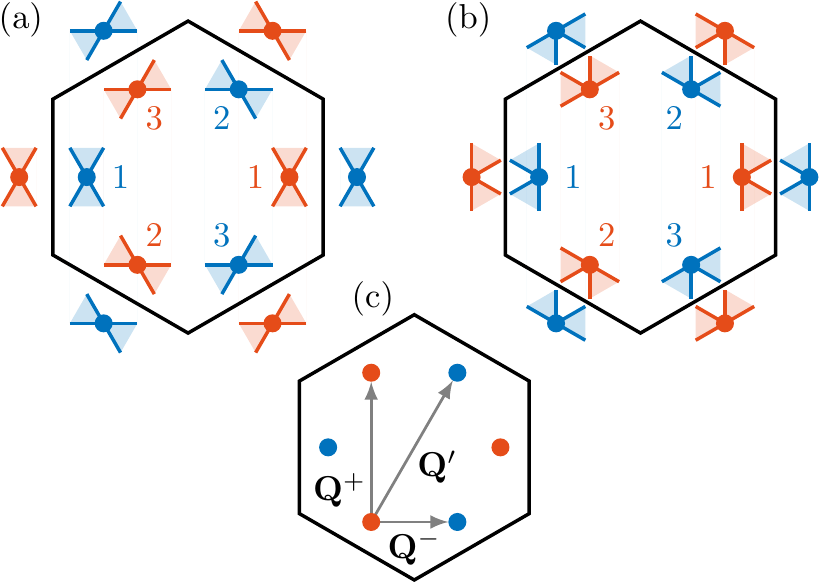}
\caption{\label{fig:os} Van Hove fermiology under orbital splitting. The originally degenerate saddle points at each $\mbf M_\a$ separate in opposite directions, with the two colors indicating different orbitals. Blank and shaded regions represent opposite sides of Fermi level. (a) In the immediate vicinity of Van Hove singularities, the $X$-like saddle points serve as more appropriate descriptions. (b) The $K$-like saddle points may apply if the regime of interest is finitely away from the Van Hove singularities. (c) Inter- and intrapatch nesting momenta.}
\end{figure}

For current graphene Moir\'e heterostructures, the low energy flat bands manifest a breakdown of orbital degeneracy \cite{model1,po18prx, model2, model3, model4,zhang19prb,zhang1809ax}. This corresponds to the breakdown of $\txt{SU}(4)$ flavor symmetry to $\txt{U}(1)_o\times\txt{SO}(4)$, where $\txt{U}(1)_o$ represents the reduced orbital symmetry, and $\txt{SO}(4)\sim\txt{SU}(2)_+\times\txt{SU}(2)_-$ is a combined spin symmetry from the two orbitals \cite{you18ax}. Two separate bands arise from the two orbitals under an intraorbital $C_{3z}$ and an interorbital $C_{2y}$ symmetries. Our interest lies in the deformation of Van Hove fermiology from the minimal model, particularly in the immediate vicinity of Van Hove singularities. A general feature is the splitting of Van Hove singularities in different orbitals at each $\mbf M_\a$, which shift into the bulk in opposite directions as illustrated in Fig.~\ref{fig:os}. The interactions now manifest sixteen inequivalent channels $g_{ij}$'s \cite{isobe18prx}, where $i,j=1,\dots,4$ denote inequivalent scatterings [Fig.~\ref{fig:bzfd}(c)] among the patches and orbitals, respectively. Momentum conservation prohibits several interactions, in particular the orbital Umklapp scattering $g_{i3}$'s. Meanwhile, the interorbital exchange $g_{i1}$'s are subleading due to the large momentum transfer at atomic scale. We therefore admit only the six eligible orbital density-density interactions $g_{14}$, $g_{22}$, $g_{24}$, $g_{32}$, $g_{42}$, and $g_{44}$ in the RG analysis, analogous to the flavor conserving setup in the minimal model. Note that the interactions $g_{12}$ and $g_{34}$ are ruled out by momentum conservation.

The shift of Van Hove singularities generally affects the Fermi surface nesting. Define the nesting momenta between intraorbital interpatch $\mbf Q^+$, interorbital interpatch $\mbf Q^-$, and interorbital intrapatch $\mbf Q'$ pairs [Fig.~\ref{fig:os}(c)], followed by the corresponding susceptibilities \cite{isobe18prx}. Cooper channel $\Pi^\txt{pp}_{\mbf 0}$ remains $\ln^2$ divergent due to an interorbital symmetry between $\pm\mbf k$. The rest channels exhibit $\ln$ or $\ln^2$ divergences depending on how the nesting is affected. Note that Ref.~\onlinecite{isobe18prx} assumes the $K$-like saddle points [Fig.~\ref{fig:os}(b)] with only interorbital nesting. Such nesting may occur near but finitely away from the Van Hove singularities. The according analysis may apply at either finite doping, finite temperature, or finite interactions. Our study focuses on the asymptotically exact weak coupling solutions, which arise primarily from the Van Hove singularities. This draws our attention to the immediate vicinity of these points. Despite the Fermi surface deformation, the $X$-like saddle points [Fig.~\ref{fig:os}(a)] still serve as more appropriate descriptions in these regions \cite{kim16nl}. We therefore adopt the patch model with $X$-like saddle points and conduct the weak coupling RG analysis. The potential nesting is similar to the interpatch nesting in the minimal model. When the nesting is present, the $\ln^2$ divergences occur in the susceptibilities $\Pi^\txt{pp}_{\mbf Q'}$ and $\Pi^\txt{ph}_{\mbf Q^\pm}$.

\subsection{Renormalization group and phase diagram}

Define the dimensionless RG time $y=\Pi^\txt{pp}_{\mbf 0}/h^\txt{pp}$ and the nesting parameters $d_0^+=d\Pi^\txt{pp}_{\mbf Q'}/d(h^\txt{pp}y)$ and $d_1^\pm=d\Pi^\txt{ph}_{\mbf Q^\pm}/d(h^\txt{pp}y)$ \cite{isobe18prx}. The RG equations read
\beeq\label{eq:rgeqtbg}\beal
\fr{dg_{14}}{dy}&=2d_1^+[g_{14}(g_{24}-g_{14})-g_{32}^2],\\
\fr{dg_{22}}{dy}&=d_1^-(g_{22}^2+g_{32}^2),\quad
\fr{dg_{24}}{dy}=d_1^+g_{24}^2,\\
\fr{dg_{32}}{dy}&=2g_{32}(d_1^-g_{22}+d_1^+g_{24}-2d_1^+g_{14})-g_{32}(g_{32}+2g_{42}),\\
\fr{dg_{42}}{dy}&=-2g_{32}^2-g_{42}^2,\quad
\fr{dg_{44}}{dy}=-d_0^+g_{44}^2,
\enal\eneq
which are similar to those with the same patch indices in the minimal model [Eq.~(\ref{eq:rgeq})]. Starting from the weak bare repulsions $g_{ij}\geq0$, the system flows to strong coupling fixed trajectories $g_{ij}=G_{ij}/(y_c-y)$ at the critical scale $y_c$. In the $i=2$ sector, both $g_{22}$ and $g_{24}$ are positive definite. $g_{32}$ is positive semidefinite, whereas $g_{14}$ and $g_{42}$ can become negative under RG. The last equation is irrelevant due to the fixed point $g_{44}=0$.

\begin{figure*}[t]
\centering
\includegraphics[scale = 1]{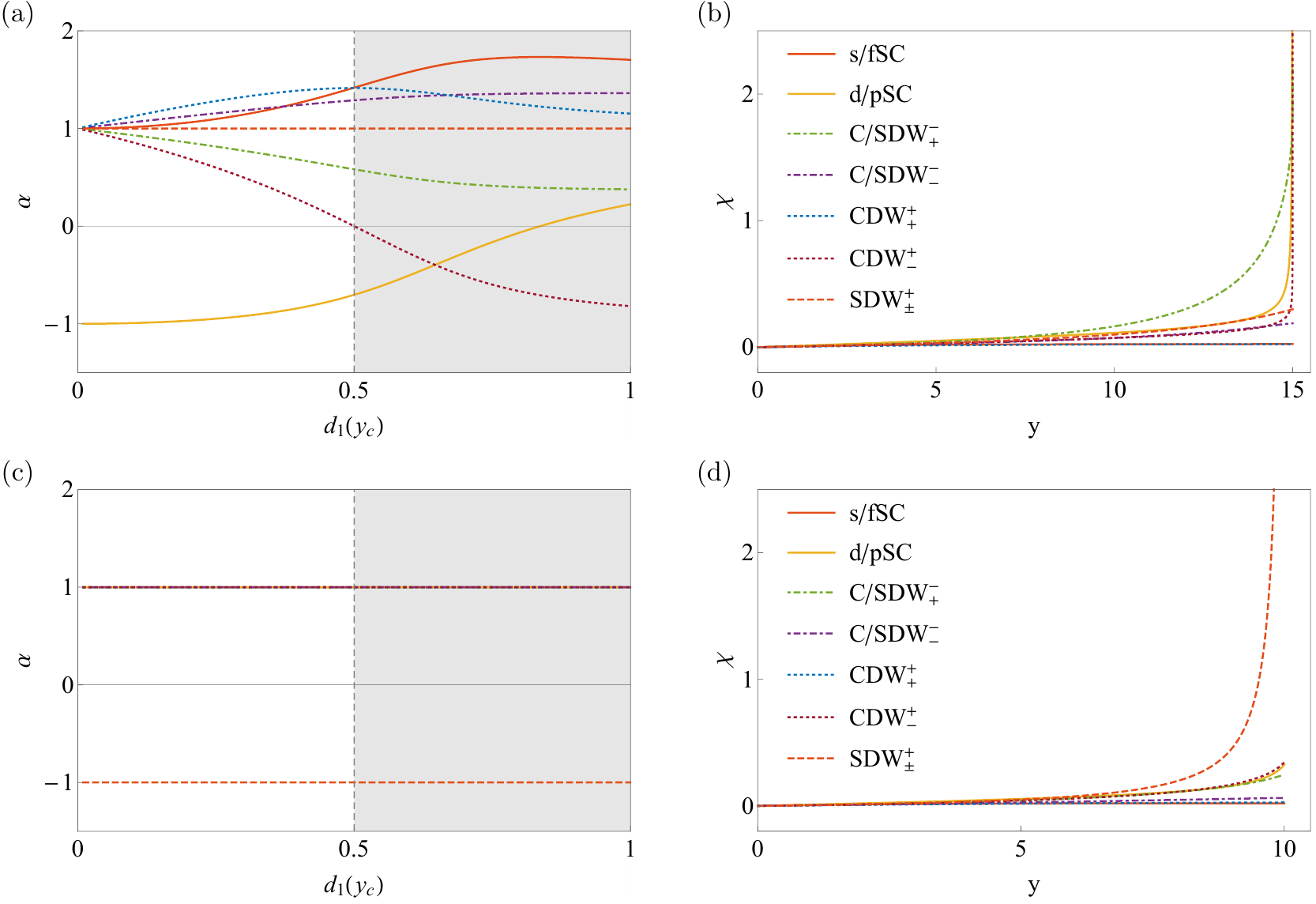}
\caption{\label{fig:tbg} Two fixed trajectories arise under orbital splitting, guided by (a)(b) inter- and (c)(d) intraorbital flavor fluctuations, respectively. We set $d_1^+=d_1^-=d_1$ in the fixed trajectory analysis (a)(c) for simplicity, and run the RG flows (b)(d) with maximal nesting $d_1^\pm=0.5$. The irrelevant $g_{44}$ and PDW$'$ are neglected. (a)(b) Starting with $g_{14}=g_{22}=g_{24}=g_{32}=g_{42}=0.1$, the interorbital flavor fluctuations guide the RG flow to the first fixed trajectory. C/SDW$^-_+$ grow first at intermediate scales. Approaching the critical scale $y_c$, the $d/p$-wave superconductivities diverge as the only leading instabilities. CDW$^+_-$ remains next leading along the fixed trajectory. (c)(d) Setting $g_{24}=0.2$ instead, the RG flow is turned to the second fixed trajectory by intraorbital flavor fluctuations. SDW$^+_\pm$ grow and become the only leading instabilities.}
\end{figure*}

The instabilities arise under the RG flow and manifest divergent susceptibilities. Unlike Ref.~\onlinecite{isobe18prx}, where the susceptibilities were determined via RPA analysis, here we determine the susceptibilities via the insertion of appropriate test vertices as in Sec.~\ref{sec:insana}. The irreducible channels are identified by the test vertex analysis. We focus on the channels receiving $\ln^2$ divergences under RG
\beeq\bega
g_\txt{$s$/$f$SC}=2g_{32}+g_{42},\quad
g_\txt{$d$/$p$SC}=-g_{32}+g_{42},\\
g_{\txt{PDW}'}=g_{44},\\
g_{\txt{SDW}^-_\pm}=g_{\txt{CDW}^-_\pm}=-(g_{22}\pm g_{32}),\\
g_{\txt{CDW}^+_\pm}=2g_{14}-g_{24}\pm2g_{32},\quad
g_{\txt{SDW}^+_\pm}=-g_{24},
\enga\eneq
where each finite momentum channel DW$^a$ is defined by the nesting momentum $\mbf Q^a$. These inter- and intraorbital channels evolve from the original superconducting and density wave channels under orbital splitting. The according orbital representations $\tau^i$'s lie in the $(O^3)^2=0,1$ orbital quantum number sectors under $\txt{U}(1)_o$ symmetry. Here $O^3$ is the third component of the orbital pseudospin. Note that the orbital and spin Pauli matrices $\tau^i$ and $\s^i$ with $\tau^0,\s^0=1$ are defined for the representations.

In the particle-particle branch, inter- and intraorbital pairings constitute zero and finite momentum channels, respectively. The interorbital pairings lead to the $s,p,d,f$-wave superconducting channels. Given the even momentum modes in the minimal model, here the even and odd modes occur from the attachment of orbital pairings $\tau^{3,0}(i\tau^2)$. For the intraorbital pairings with $\tau^{1,2}(i\tau^2)$, the pair density wave PDW$'$ does not develop due to the fixed point $g_{44}=0$. Meanwhile, the particle-hole branch also manifests various inter- and intraorbital channels. The interorbital C/SDW$^-_\pm$'s manifest the test vertices $\D_{\mbf Q^-\tau\pm}(\psi_{\b\tau}^\dag\psi_{\a\tau'}\pm\psi_{\a\tau}^\dag\psi_{\b\tau'})$, where $\tau\neq\tau'$ are the orbital indices. These channels evolve from the original interorbital FDWs with $\tau^{1,2}$ representations. The spin singlet and triplet representations $\s^0,\bsb\s$ determine whether the states are CDW$^-_\pm$ or SDW$^-_\pm$. For the intraorbital sector, the test vertices $\D_{\mbf Q^+\tau\pm}(\psi_{\b\tau}^\dag\psi_{\a\tau}\pm\psi_{\a\tau'}^\dag\psi_{\b\tau'})$ define C/SDW$^+_\pm$'s. These channels evolve from the original CDWs and intraorbital FDWs with $\tau^{0,3}$ representations. SDW$^+_\pm$ develop from the intraorbital FDWs with $\bsb\s$ representations. CDW$^+_\pm$, on the other hand, arise from CDWs and intraorbital FDWs with $\s^0$ representation.

The test vertex analysis conclude all potential instability channels near Van Hove filling. Note the inclusion of density wave channels beyond $s$-wave \cite{nayak00prb}, which are ignored in the previous analyses \cite{isobe18prx,you18ax}. These high angular momentum channels resemble the imaginary density waves in the minimal model. With the imaginary CDW proven dominant in the minimal model, we will show that the relevance still holds under orbital splitting.

Analyzing the RG equations Eq.~(\ref{eq:rgeqtbg}), we identify two fixed trajectories and the according instabilities (Fig.~\ref{fig:tbg}). Significantly, the distinction inbetween features the competition between inter- and intraorbital flavor fluctuations, corresponding to the $(O^3)^2=1,0$ sectors, respectively. A quantitative analysis concerns whether $g_{22}$ or $g_{24}$ diverges first under RG.

When $g_{22}$ wins, the interorbital flavor fluctuations guide the RG flow to the first fixed trajectory [Figs.~\ref{fig:tbg}(a)(b)]. This RG flow is similar to that in the minimal model. At intermediate scales, leading C/SDW$^-_+$ manifest the interorbital flavor fluctuations, consistent with previous RPA probe of instabilities \cite{isobe18prx,you18ax}. As RG runs further, the fixed trajectory instabilities receive enhancement and become divergent. The critical interactions $G_{32}>G_{22}>0$, $G_{14},G_{42}<0$, and $G_{24}=0$ result in a similar phase diagram to Fig.~\ref{fig:rg}(b), where the $d/p$-wave superconductivities and CDW$^+_-$ are leading. However, the transition nesting now sits beyond the available regime $d_1^c>1/2$. This corresponds to a suppression of CDW$^+_-$, which occurs since the internal fermion loop loses the enhancement from intraorbital flavor fluctuations. Therefore, the $d/p$-wave superconductivities are the only leading instabilities along this fixed trajectory. Note that CDW$^+_-$ is still next leading, indicating its significance near Van Hove filling.

A different fixed trajectory arises when $g_{24}$ diverges prior to $g_{22}$ [Fig.~\ref{fig:tbg}(c)(d)]. Since the intraorbital flavor fluctuations win, the growth of SDW$^+_\pm$ at intermediate scales is expected.  However, unlike the RG flows discussed previously, SDW$^+_\pm$ now retain the leading role along the whole trajectory. This is confirmed by the fixed trajectory analysis, where only two negative susceptibility exponents $\a_{\txt{SDW}^+_\pm}=-1$ arise from the critical interactions $G_{24}=2G_{14}>0$ and $G_{22}=G_{32}=G_{42}=0$. The monopoly occurs since $g_{34}$ is prohibited by momentum conservation. Without Umklapp scattering in the intraorbital sector, intraorbital flavor fluctuations do not share its strength to any other channel. Therefore, SDW$^+_\pm$ keep growing under RG and become the only leading instabilities along this fixed trajectory.

A phase diagram can be determined according to the RG analyses. The correlated phases depend on the doping, orbital splitting, and competition between inter- and intraorbital flavor fluctuations. When the orbital splitting is absent, the imaginary CDW dominates near Van hove filling. The $d/p$-wave superconductivities and SDW$^+_\pm$ take over under doping and orbital splitting. A caveat arises when comparing Fig.~\ref{fig:rg}(b) and Fig.~\ref{fig:tbg}(a), where the transition nesting $d_1^c$ experiences a discontinuous jump. This suggests that the imaginary CDW may only be leading in the minimal model. We note that this observation is actually an artifact of our model setup and the analysis thereof. Our analysis has adopted a patch model near Van Hove singularities and admitted only $\ln^2$ divergences in RG. The asymptotically exact solutions are extracted in the weak coupling limit $g_i\rar0$. In the practical systems, the interactions are away from the weak coupling limit. These finite interactions can drag in the neglected ingredients, including the Fermi surface sectors outside the patch model and the $\ln$ divergences in RG. The discontinuous jumps between fixed trajectories from different setups are then smeared out, leading to continuous transitions or crossovers between according correlated phases. A finite regime of CDW$^+_-$ upon orbital splitting is thus expected, which gives way to the $d/p$-wave superconductivities and SDW$^+_\pm$ at certain splitting scale. On the other hand, the transition between the later two phases manifests an `unstable' fixed trajectory, where only one positive eigenvalue is uncovered in the linearized RG equations (derived following Appendix~\ref{app:ft}). The susceptibility exponents are identical to those along the first fixed trajectory in Fig.~\ref{fig:tbg}(a), but with different leading instabilities $\a_{\txt{SDW}^+_\pm}=-1$. This fixed trajectory may also describe the transition between CDW$^+_-$ and SDW$^+_\pm$.

Our analysis applies when the nesting remains similar to the minimal model under orbital splitting. This setup captures all potential leading instabilities which receive $\ln^2$ divergences under RG. However, the results may become unstable when the nesting is destroyed by the fermiology deformation under orbital splitting. At low nesting regime, the $d/p$-wave superconductivities are predominantly the leading instabilities in the weak coupling limit. If the nesting is completely destroyed, the system may return to the normal metal. On the other hand, if the regime of interest is shifted by finite doping, temperature, or interaction strength, the fermiology may acquire an appropriate description by the $K$-like saddle points [Fig.~\ref{fig:os}(b)]. The nesting now occurs only in the interorbital sector. With $\ln^2$ divergent $\Pi^\txt{pp}_{\mbf Q^+}$ and $\Pi^\txt{ph}_{\mbf Q^-,\mbf Q'}$, a different set of RG equations is determined \cite{isobe18prx}. Interorbital flavor fluctuations guide the RG flow to a unique fixed trajectory. While C/SDW$^-_+$ grows at intermediate scales, the $d/p$-wave superconductivities dominates universally along the fixed trajectory. CDW$^+_-$ does not receive $\ln^2$ divergence and becomes irrelevant. These observations are consistent with the RPA probe of instabilities in previous analyses \cite{isobe18prx,you18ax}.

\subsection{Correlated phases}

The leading instabilities manifest similar features to the corresponding channels in the minimal model. Subtle differences occur due to the incommensuration away from $\mbf M_\a$'s. Various potential configurations in the $d/p$-wave superconducting channels have been discussed \cite{you18ax}. To maximize the ordering energy, the chiral $d+id$ and $p+ip$ superconductivities are expected to dominate, which manifest $[\tau^3(i\tau^2)][\s^0(i\s^2)]$ and $[\tau^0(i\tau^2)][\bsb\s(i\s^2)]$ representations. The quantitative justification requires a complete Landau-Ginzburg analysis, which is an interesting problem to explore for future work. Note that a spin triplet pairing is manifested in the $p+ip$ channel, leading to a suppression of critical temperature [Eq.~(\ref{eq:supptc})].

The density wave instabilities result from CDW and intraorbital FDW channels in the minimal model. For CDW$^+_-$, an interorbital minus sign $\epvl{\psi_{\a\tau}^\dag\psi_{\b\tau}}=-\epvl{\psi_{\a\tau'}^\dag\psi_{\b\tau'}}^*$ occurs between channels with $\tau^0\pm\tau^3$ representations. This implies a manifold with degenerate imaginary CDW ($\tau^0\s^0$) and real intraorbital spin singlet FDW ($\tau^3\s^0$) from the minimal model. The imaginary CDW has been identified as a QAHI with charge loop current. The real intraorbital spin singlet FDW, on the other hand, exhibits a half metal from the $s$-wave uniaxial $3Q$ orbital density wave (ODW) with $\tau^3$ orbital modulation \cite{nandkishore12prl,venderbos16prb2}. Both states manifest real space patterns with incommensurate periods. Note that the incommensuration of nesting momenta leads to a mixture with other angular momentum modes \cite{nayak00prb}. When the deviation from $\mbf M_\a$'s is small, charge modulation and orbital loop current occur perturbatively. The resulting states generally break time reversal symmetry as the parent QAHI and half metal \cite{venderbos16prb2}. Since QAHI exhibits a full gap while the half metal does not, we expect QAHI to be the leading state within this manifold. A Landau-Ginzburg analysis is required for the rigorous justification, which is an interesting problem for future work. Meanwhile, the degenerate SDW$^+_\pm$ channels manifest a larger manifold. We discuss first the $\tau_0$ channels. The SDW$^+_+$ with $s$-wave pattern is smoothly connected to the SDW states discussed in Refs.~\onlinecite{martin08prl,nandkishore12prl,Fernandes,chern12prl,venderbos16prb2}. At zero temperature, 3Q non-coplanar order is expected, realizing a Chern insulator \cite{martin08prl}. SDW$^+_-$ is a state of a sort discussed in Ref.~\onlinecite{venderbos16prb2}. If uniaxial $3Q$ order develops then a quantum spin Hall insulator (QSHI) is realized. If $3Q$ non-coplanar order develops then a spin-locked Dirac semimetal is realized. We expect QSHI to be favored since it manifests a full gap absent in the Dirac semimetal. To quantitatively determine the nature of the ordering requires a Landau-Ginzburg analysis, which is beyond the scope of this work. Finally, the $\tau^3$ channels ($\tau^0\bsb\s$) contain additional density wave possibilities which are interesting problems to explore for future work. Note that the SDW$^+_\pm$ channels exhibit the breakdown of continuous $\txt{SU}(2)$ spin symmetry. A suppressed critical temperature similar to Eq.~(\ref{eq:supptc}) is thus expected.

\section{Square lattice}

\begin{figure}[t]
\centering
\includegraphics[scale = 1]{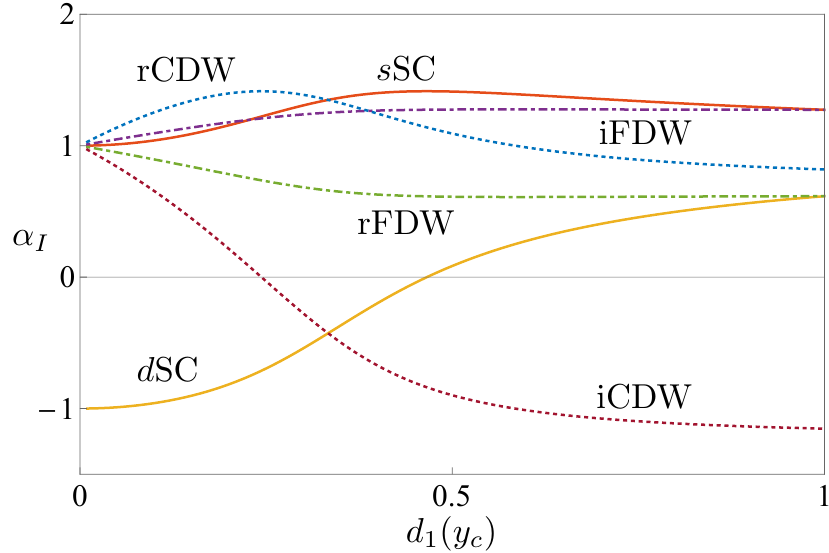}
\caption{\label{fig:slf4se} Susceptibility exponents in all channels for $N_f=4$ on the square lattice $N_p=2$.}
\end{figure}

While we have focused on hexagonal lattices for our analysis, our RG equations also works for the square lattice systems if we set $N_p=2$. With spin degeneracy only $N_f=2$ this maps onto the problem studied by Refs.~\onlinecite{dzyaloshinski, schulz, furukawa}, and exhibits a `conventional high-$T_c$' phase diagram with an insulating antiferromagnetic phase flanked by nodal superconducting domes. Things become different when $N_f>2$. With the flavor number set as $N_f=4$, the susceptibility exponents are evaluated and presented in Fig.~\ref{fig:slf4se}. As in the previous studies with only spin degeneracy \cite{schulz,dzyaloshinski,furukawa}, the (nodal) $d$-wave superconductivity is dominant at low nesting regime, and becomes degenerate with the real FDW state at perfect nesting $d_1=1$ to leading order $O(\ln^2)$. However, the imaginary CDW overcomes these degenerate channels and becomes the leading instability in the high nesting regime. The density wave order develops at a single momentum, since there is only one nesting momentum on square lattice. This loop current state is a gapless $d$-wave staggered flux state \cite{affleck88prb,chakravarty01prb,honerkamp04prl}. Despite the breakdown of time reversal and translational symmetries, an effective time reversal symmetry assisted by a translation is present \cite{venderbos16prb}. Notice that this staggered flux state is different from the gapped Varma loop current state \cite{varma99prl}, where the translational symmetry is preserved.

A transition between the $d$-wave superconductivity and the staggered flux state is universal for square lattice with extra flavors $N_f>2$. In the doping phase diagram, a staggered flux state is observed near the Van Hove filling, while $d$-wave superconductivity takes over upon doping away. This transition can be regarded as the `nodal' and `nonchiral' version of the transition observed on hexagonal lattices. When the flavor number increases, the transition point moves toward low nesting regime, implying an expansion of the staggered flux state in the phase diagram.

\section{Discussion}
\label{sec:concl}

We have analyzed repulsively interacting fermions on hexagonal lattices near Van Hove filling with an $\txt{SU}(N_f)$ flavor degeneracy ($N_f\geq4$), using a combination of parquet RG and Landau-Ginzburg analysis. We have determined that the leading instability at Van Hove filling is to a fully gapped chiral insulator exhibiting quantum anomalous Hall effect, which gives way upon doping to a chiral $d+id$ superconducting phase. The phase diagram consists of an insulating phase flanked by superconducting domes, reminiscent of high-$T_c$ materials (which is also manifested in experiments \cite{tbg1, tbg2, yankowitz19sci}), however both the insulating and superconducting phases are {\it chiral}. The trigger of these phases by the mediation of flavor fluctuations has also been identified. We have further examined the effects of orbital splitting, which usually occurs in the graphene Moir\'e heterostructures. With the knowledge gained from the minimal model, we have established a transparent picture of how the correlated phases arise from the competition between inter- and intraorbital flavor fluctuations. The robustness of chiral phases in the minimal model against orbital splitting has been examined. Additional spin density waves have also been uncovered. While our work has focused on the asymptotically exact weak coupling solutions, the moderate coupling results can also be recognized. The analysis bridges the minimal model and the practical Moir\'e band structures. Useful hints may be provided for the interpretation of experimentally observed correlated phases under various circumstances.

Note that the Chern insulator we uncover in the minimal model is significantly different from those proposed in previous studies of graphene Moir\'e heterostructures \cite{PRL,zhang19prb,zhang1809ax,liudai18prb,xie18ax}. The Chern insulators in these other works arise from magnetic ordering in the spin-valley sector. In contrast, our Chern insulator does not exhibit any magnetic order. The quantum anomalous Hall effect is triggered instead by the local fluxes in the ground state, closer to the original proposal by Haldane \cite{Haldane}. Subsequent to the first posting of our work, an experiment reported the anomalous Hall effect in the twisted bilayer graphene at angle $\t\apx1.17\dge$, which is aligned with a hexagonal boron nitride \cite{sharpe19ax}. Although the measured Hall conductivity is not strictly quantized, the observation still provides crucial signals of nontrivial topological bands. Theoretical works on this system again propose a Chern insulator with magnetic order \cite{bultinck19ax,zhang19ax}. Another experiment reported the loop current states at larger angles, such as $\t\apx3.89\dge$, in the presence of external magnetic field \cite{weckbecker19ax}. Our work may provide some indications on how these recent experimental observations can be interpreted.

It should be stressed that our model is extremely simplified and neglects many experimental details. It is also not clear whether the experiments in Refs.~\onlinecite{tbg1, tbg2,3lg,yankowitz19sci, kerelsky18ax} are really well described by weak coupling. However, the simplicity of the model helps expose the essential physics, and reveals that hexagonal lattices with enhanced flavor degeneracy (which may arise naturally in Moir\'e heterostructures) should provide an ideal platform for realizing tunable chiral insulating and superconducting phases. Whether the correlated phases observed in present or future Moir\'e heterostructures are actually chiral should be straightforwardly testable in experiment.

\section{Acknowledgments}
We acknowledge Sankar Das Sarma and Andrey V. Chubukov for valuable feedback on the manuscript. This research was sponsored by the Army Research Office and was accomplished under Grant No. W911NF-17-1-0482. The views and conclusions contained in this document are those of the authors and should not be interpreted as representing the official policies, either expressed or implied, of the Army Research Office or the U.S. Government. The U.S. Government is authorized to reproduce and distribute reprints for Government purposes notwithstanding any copyright notation herein.


\appendix

\section{Patch models}
\label{app:patch}

On a two dimensional lattice, the dispersion energy exhibits saddle points at the Van Hove fillings. A patch model can be defined for the low energy theory, where the Fermi surface is approximated by the patches at the saddle points. Each patch is defined by a momentum cutoff with respect to an energy scale $\L$. We introduce the patch models on the square and hexagonal lattices. Only nearest neighbor hoppings (with hopping constant $t>0$) are included, which is rationalized by the large unit cell in the study of Moir\'e superlattice models.

\subsection{Square lattice}

For the square lattice, the Brillouin zone is a square with corners $(\pm\pi,\pm\pi)$. At Van Hove filling, the Fermi surface is a smaller square defined by the inequivalent corners $\mbf M_1=(\pi,0)$ and $\mbf M_2=(0,\pi)$ on the Brillouin zone boundary. These corners are the saddle points of dispersion energy
\beeq
\label{eq:engsq}
\beal
\xi_{\mbf M_1+\mbf k}&\apx t\lf(-k_x^2+k_y^2\ri),\\
\xi_{\mbf M_2+\mbf k}&\apx t\lf(k_x^2-k_y^2\ri),
\enal\eneq
where $\mbf k$ represents the infinitesimal deviation from the saddle points.

\subsection{Hexagonal lattices}

The class of hexagonal lattices includes triangular and honeycomb lattices with hexagonal Brillouin zones. For the triangular lattice, the corners of the Brillouin zone are $(0,\pm4\pi/3)$ and $(\pm2\pi/\sqrt3,\pm2\pi/3)$. At Van Hove filling, the Fermi surface is a hexagon inscribed in the Brillouin zone. The inequivalent corner saddle points $\mbf M_1=(-2\pi/\sqrt3,0)$ and $\mbf M_{2,3}=(\pi/\sqrt3,\pm\pi)$ exhibit the dispersion energies
\beeq
\label{eq:engtr}
\beal
\xi_{\mbf M_1+\mbf k}&\apx \fr{t}{2}\lf(-3k_x^2+k_y^2\ri),\\
\xi_{\mbf M_{2,3}+\mbf k}&\apx tk_y\lf(\mp\sqrt3k_x-k_y\ri).
\enal\eneq
Similar features are manifested on the honeycomb lattice. However, Van Hove filling exists in both electron and hole doping regimes. We only focus on the hole doping regime, with the analyses for the electron doping regime being similar. The Brillouin zone is defined by the six corner Dirac points $(0,\pm4\pi/3\sqrt3)$ and $(\pm2\pi/3,\pm2\pi/3\sqrt3)$. At Van Hove filling, the hexagonal Fermi surface exhibits inequivalent corner saddle points $\mbf M_1=(-2\pi/3,0)$ and $\mbf M_{2,3}=(\pi/3,\pm\pi/\sqrt3)$, with the dispersion energies
\beeq
\label{eq:enghc}
\beal
\xi_{\mbf M_1+\mbf k}&\apx \fr{3t}{4}\lf(-3k_x^2+k_y^2\ri),\\
\xi_{\mbf M_{2,3}+\mbf k}&\apx \fr{3t}{2}k_y\lf(\mp\sqrt3k_x-k_y\ri).
\enal\eneq

\section{Susceptibilities and maximal nesting at Van Hove filling}
\label{app:nesting}

Near Van Hove filling, the intrapatch particle-particle and interpatch particle-hole susceptibilities
\beeq\beal
\Pi^\txt{pp}_{\mbf 0}&=T\sum_n\intv{k}G_{\mbf k\o_n}G_{(-\mbf k)(-\o_n)},\\
\Pi^\txt{ph}_{\mbf Q}&=-T\sum_n\intv{k}G_{\mbf k\o_n}G_{(\mbf k+\mbf Q)\o_n}
\enal\eneq
can acquire $\ln^2$ divergence. The fermionic Matsubara frequency $\o_n=(2n+1)\pi T$ is summed within the energy cutoff $[-\L,\L]$, while the momentum integral $\intv{k}=\int d^2k/(2\pi)^2$ is conducted within the patches. Since the Fermi surface nesting only affects particle-hole susceptibility $\Pi^\txt{ph}_{\mbf Q}$, a quantitative measure of nesting degree is provided by the nesting parameter $d_1=d\Pi^\txt{ph}_{\mbf0}/d\Pi^\txt{pp}_{\mbf Q}$.

The range of available nesting parameters is an important issue for the phase transition analysis. If a transition point $d_1^c$ between instabilities existed within this range, a phase transition could occur. Notice that the nesting parameter is positive semidefinite $d_1\geq0$. The maximum, on the other hand, is determined by the value $d_1^\txt{max}$ at Van Hove filling. In the weak coupling and low temperature $T\ll t$ limit, this maximum $d_1^\txt{max}=h^\txt{ph}/h^\txt{pp}$ is determined by the prefactors of the two susceptibilities at leading order $O(\ln^2)$. The geometry of Fermi surface affects this value significantly, as will be elucidated in the remaining part of this section.

\subsection{Square lattice}

The two susceptibilities $\Pi^\txt{pp}_{\mbf 0}$ and $\Pi^\txt{ph}_{\mbf Q}$ are identical on square lattice. With the dispersion energies Eq.~(\ref{eq:engsq}), both $\Pi^\txt{pp}_{\mbf 0}$ and $\Pi^\txt{ph}_{\mbf Q}$ exhibit the form
\beeq
\Pi
=T\sum_n\intv{k}\fr{1}{[i\o_n+t(k_x^2-k_y^2)][-i\o_n+t(k_x^2-k_y^2)]}.
\eneq
The result $d_1^\txt{max}=1$ indicates that the patches can gain access to the perfect nesting at Van Hove filling.

An evaluation of the $\ln^2$ divergence is useful for later analysis of hexagonal lattices. Rewrite the momentum integral in terms of the parameters $a_\pm=\sqrt t(\pm k_x+k_y)$
\beeq
\label{eq:ppssqa}
\Pi
=hT\sum_n\int_{-\sqrt\L}^{\sqrt\L} da_+da_-\fr{1}{(i\o_n-a_+a_-)(-i\o_n-a_+a_-)}
\eneq
with the prefactor $h=(8\pi^2t)^{-1}$ manifesting the characteristic density of states. The integral domain defines the setup of square patches. After a reparametrization $x=a_+a_-$
\beeq
\Pi
=hT\sum_n\int_{-\sqrt\L}^{\sqrt\L} \fr{da_+}{a_+}\int_{-\sqrt\L a_+}^{\sqrt\L a_+}dx\fr{1}{(-i\o_n+x)(i\o_n+x)},
\eneq
an infrared (IR) cutoff $|a_+|\geq|\o_n|/\sqrt\L$ is imposed for the $a_+$ integral. The regime below this cutoff is neglected since it only provides a $\ln$ divergence. Approximating the $x$ integral by integrating over the whole real axis, the Cauchy integral formula provides
\beeq
\Pi\apx hT\sum_n\fr{2\pi}{|\o_n|}\int_{|\o_n|/\sqrt\L}^{\sqrt\L}\fr{da_+}{a_+}.
\eneq
A factor $\txt{sgn}(a_+)$ arises from the integral and cancels the sign of $a_+$ in the denominator. After the $a_+$ integral
\beeq
\label{eq:afteraint}
\Pi\apx hT\sum_n\fr{2\pi}{|\o_n|}\ln\fr{\L}{|\o_n|},
\eneq
the Matsubara frequency summation
\beeq
T\sum_n\rar2\int_{2\pi T}^{\L}\fr{d\o}{2\pi}
\eneq
reveals the $\ln^2$ divergence of the susceptibility
\beeq
\label{eq:susln2}
\Pi\apx h\ln^2\fr{\L}{T}.
\eneq
The prefactor $h^{\txt{pp}/\txt{ph}}=h$ is consistent with Ref.~\onlinecite{furukawa}, giving the maximal nesting $d_1^\txt{max}=1$.

\subsection{Hexagonal lattices}

We proceed to conduct the same calculations on hexagonal lattices. Our analysis focuses on the triangular lattice with dispersion energies Eq.~(\ref{eq:engtr}). The results are also applicable to the honeycomb lattice.

\subsubsection{Particle-particle susceptibility}

We start by determining $\Pi^\txt{pp}_{\mbf0}$ at $\mbf M_1$. With the reparametrization $a_\pm=\sqrt{t/2}(\pm\sqrt3 k_x+k_y)$, the integral Eq.~(\ref{eq:ppssqa}) is again obtained, but the prefactor is now $h^\txt{pp}=h$ with $h=(4\sqrt3\pi^2t)^{-1}$. The integral domain defines a diamond patch centered at $\mbf M_1$. A $\ln^2$ divergence Eq.~(\ref{eq:susln2}) is derived as on square lattice.

The calculations at the other saddle points $\mbf M_2$ and $\mbf M_3$ are also conducted. With the integral domain unchanged, the alignments of diamond patches with the saddle point structures become different. Despite such difference, the calculations again derive the same result. This feature indicates the universality of IR properties at Van Hove filling. As long as the whole saddle point structure is covered, a deformation of patch shape only leads to a change in the subleading terms.

\subsubsection{Particle-hole susceptibility}

We turn to calculate $\Pi^\txt{ph}_{\mbf Q}$ for $\mbf Q_3=\mbf M_2-\mbf M_1$
\beeq\beal
\Pi^\txt{ph}_{\mbf Q}
&=-hT\sum_n\int_{-\sqrt\L}^{\sqrt\L}da_+da_-\times\\
&\quad\fr{1}{(i\o_n-a_+a_-)[i\o_n+a_+(a_++a_-)]}.
\enal\eneq
The reparametrization $x=a_+a_-$ leads to
\beeq\beal
\Pi^\txt{ph}_{\mbf Q}
&=hT\sum_n\int_{-\sqrt\L}^{\sqrt\L}\fr{da_+}{a_+}\int_{-\sqrt\L a_+}^{\sqrt\L a_+}dx\times\\
&\quad\fr{1}{(-i\o_n+x)(i\o_n+a_+^2+x)},
\enal\eneq
and the IR cutoff $|a_+|\geq|\o_n|/\sqrt\L$ is again imposed. Integrate $x$ over the whole real axis
\beeq
\label{eq:phsaint}
\Pi^\txt{ph}_{\mbf Q}
\apx hT\sum_n4\pi\int_{|\o_n|/\sqrt\L}^{\sqrt\L}\fr{da_+}{a_+}\fr{i\txt{sgn}(\o_n)(-2i\o_n+a_+^2)}{4\o_n^2+a_+^4}.
\eneq
Since the imaginary part is odd in Matsubara frequency, only the real part survives the Matsubara frequency summation. Taking $y=a_+^4$, the susceptibility becomes
\beeq
\Pi^\txt{ph}_{\mbf Q}
\apx hT\sum_n2\pi|\o_n|\int_{\o_n^4/\L^2}^{\L^2}\fr{dy}{y}\fr{1}{4\o_n^2+y}.
\eneq
Decompose the integrand
\beeq
\fr{1}{y(4\o_n^2+y)}=\fr{1}{4\o_n^2}\lf(\fr{1}{y}-\fr{1}{4\o_n^2+y}\ri)
\eneq
and evaluate the integral, where an approximation $4\o_n^2+\o_n^4/\L^2\apx4\o_n^2$ is imposed in the second term. We arrive at the result Eq.~(\ref{eq:afteraint}) with a different prefactor $h^\txt{ph}=h/2$. The Matsubara frequency summation then uncovers the $\ln^2$ divergence of the susceptibility Eq.~(\ref{eq:susln2}). This result is also derived for $\mbf Q_1$ and $\mbf Q_2$.

\subsubsection{Nesting parameter}

Having derived the $\ln^2$ divergences for the susceptibilities, we identify the maximal nesting parameter $d_1^\txt{max}=1/2$. The result indicates that the patches on hexagonal lattices do not enjoy the perfect nesting of Fermi surface. However, the finite nesting degree still provides a chance for the density waves to arise.

\section{Fixed trajectories}
\label{app:ft}

In this section, we analyze the parquet renormalization group (RG) equations provided in the main text. Consider the setup of bare repulsive interactions $g_1,g_2,g_3,g_4\geq0$. When $d_1(y)>0$, $g_2$ increases monotonically under the RG flow, indicating the strong coupling fixed trajectories. The divergence occurs at a certain scale $y=y_c$. This feature suggests a simplification of RG equations by taking $g_2$ as a new RG time \cite{nandkishore12np}. Define the fixed point nesting parameter $d_1=d_1(y_c)$ and rewrite the RG equations in terms of $x_i=g_i/g_2$
\beeq\beal
\fr{dx_1}{d\ln g_2}&=-x_1+\fr{x_1(2-N_fx_1)+(2-N_f)x_3^2}{1+x_3^2},\\
\fr{dx_3}{d\ln g_2}&=-x_3\\
&+\fr{2d_1x_3[2-(N_f-1)x_1]-x_3[(N_p-2)x_3+2x_4]}{d_1(1+x_3^2)},\\
\fr{dx_4}{d\ln g_2}&=-x_4+\fr{-(N_p-1)x_3^2-x_4^2}{d_1(1+x_3^2)}.
\enal\eneq
The fixed point solutions $x_i=x_i^*$ are determined by the static condition $dx_i/d\ln g_2=0$.

The equations provide various sets of solutions, representing various fixed points of the RG flow. However, there are constraints on the eligible fixed points, including the real condition $g_i\in\mbb R$ and the positive semidefinite condition $g_i\geq0$. For $N_f=2$, both $g_1$ and $g_3$ should be positive semidefinite. However, when $N_f>2$, $g_1$ can become negative due to a $-g_3^2$ term. The eligible fixed points should also be stable. To examine the stability of a fixed point, we derive the linearized RG equations for the infinitesimal displacements $\d x_i=x_i-x_i^*$
\beeq
\fr{d}{d\ln g_2}\lf(\bear{c}\d x_1\\\d x_3\\\d x_4\enar\ri)=\lf(\bear{ccc}M_{11}&M_{13}&M_{14}\\M_{31}&M_{33}&M_{34}\\M_{41}&M_{43}&M_{44}\enar\ri)\lf(\bear{c}\d x_1\\\d x_3\\\d x_4\enar\ri).
\eneq
The matrix elements $M_{ij}=(\p f_i/\p x_j)_{x_k=x_k^*}$ are determined from the RG equations $dx_i/d\ln g_2=f_i(x_1,x_3,x_4)$. Each eigenvalue $\l$ of the matrix $M=(M_{ij})$ indicates the flow direction along the corresponding eigenvector in phase space. The interactions flow toward the fixed point when $\l<0$, and vice versa. A fixed point is stable only when all of the eigenvalues are negative.

Analyzing the RG equations presented in the main text in this manner, we find that for $N_p=3$ there is only one stable fixed trajectory consistent with the constraints. Thus, the system flows to this fixed trajectory for generic initial repulsive interactions. The divergence of interactions can be determined as follows. Near the critical scale $y\apx y_c$, the RG equation for $g_2$ takes the form
\beeq
\fr{dg_2}{dy}=d_1(1+x_3^2)g_2^2,
\eneq
where the approximations $d_1\apx d_1(y_c)$ and $x_3\apx x_3(y_c)$ are adopted. The solution implies the critical scaling Eq.~(\ref{eq:gans}), where the critical interactions $G_i=x_i/[d_1(1+x_3^2)]$ characterize the divergence of $g_i$'s in a quantitative way. Notice that the analysis presented here is equivalent to the direct adoption of critical scaling ansatz Eq.~(\ref{eq:gans}) described in the main text.

\section{Test vertex analysis}
\label{app:inst}

In the instability analysis, the test vertices coupled to various fermion bilinears are introduced \cite{chubukov08prb,cvetkovic12prb,chubukov16prx}. By analyzing the flow of test vertices, the irreducible channels can be uncovered. We identify the superconducting and density wave channels in this section. The interaction in each channel is obtained by deriving the corresponding flow equation Eq.~(\ref{eq:tvcorrection}).

The superconducting channels exhibit intrapatch particle-particle pairings $\psi_{\a\s}\psi_{\a\s'}$ between different flavors $\s>\s'$. Accordingly, the test vertices are added as
\beeq
\d H=\sum_{\a=1}^{N_p}(\D_\a\psi_{\a\s}^\dag\psi_{\a\s'}^\dag+\D_\a^*\psi_{\a\s}\psi_{\a\s'}).
\eneq
At the $\ln^2$ order, the available corrections to the test vertices manifest the equation
\beeq
\fr{d\D_\a}{dy}=-g_4\D_\a-g_3\sum_{\b\neq\a}\D_\b,
\eneq
with a matrix representation
\beeq
\fr{d}{dy}\lf(\bear{c}\D_1\\\D_2\\\vdots\\\D_{N_p}\enar\ri)
=-\lf(\bear{cccc}g_4&g_3&\dots&g_3\\g_3&g_4&\ddots&\vdots\\\vdots&\ddots&\ddots&g_3\\g_3&\dots&g_3&g_4\\\enar\ri)
\lf(\bear{c}\D_1\\\D_2\\\vdots\\\D_{N_p}\enar\ri).
\eneq
The irreducible pairing channels are identified with the eigenstates of this equation. For the square and hexagonal lattices, a single $s$-wave and $N_p-1$ degenerate $d$-wave channels are uncovered. The effective interactions in Eq.~(\ref{eq:tvcorrection}) correspond to the eigenvalues
\beeq
g_s=(N_p-1)g_3+g_4,\quad
g_d=-g_3+g_4.
\eneq

The density wave channels can be evaluated with similar procedure \cite{chubukov08prb,chubukov16prx}. Here the fermion bilinears manifest interpatch particle-hole pairing $\psi_{\a\s}^\dag\psi_{\b\s'}$ with $\a>\b$. Notice the choice of permutative patch convention $1>3$ so that the corresponding nesting momentum takes the form $\mbf Q_2=\mbf M_1-\mbf M_3$. Either the same $\s=\s'$ or different $\s\neq\s'$ flavors are paired within each pairing.

We first analyze the density wave channels with uniform flavor pairings
\beeq
\d H=\sum_\s(\D_\s\psi_{\b\s}^\dag\psi_{\a\s}+\D_\s^*\psi_{\a\s}^\dag\psi_{\b\s}).
\eneq
The corrections to the test vertices are described by a differential equation
\beeq
\fr{d}{dy}\lf(\bear{c}\vec\D\\\vec\D^*\enar\ri)=d_1\lf(\bear{cccccc}M_\txt{d}&M_\txt{o}\\M_\txt{o}&M_\txt{d}\enar\ri)\lf(\bear{c}\vec\D\\\vec\D^*\enar\ri),
\eneq
where the real and imaginary test vertex vectors $\vec\D^{(*)}=(\D_{\s_1}^{(*)},\dots,\D_{\s_{N_f}}^{(*)})^T$ are defined. The diagonal and offdiagonal blocks of the correction matrix $M$ take the form
\beeq
\beal
M_\txt{d}&=\lf(\bear{cccc}g_2-g_1&-g_1&\dots&-g_1\\-g_1&g_2-g_1&\ddots&\vdots\\\vdots&\ddots&\ddots&-g_1\\-g_1&\dots&-g_1&g_2-g_1\\\enar\ri)
,\\
M_\txt{o}&=\lf(\bear{cccc}0&-g_3&\dots&-g_3\\-g_3&0&\ddots&\vdots\\\vdots&\ddots&\ddots&-g_3\\-g_3&\dots&-g_3&0\\\enar\ri).
\enal
\eneq
To diagonalize the correction matrix $M$, we work in a simultaneous eigenbasis for both the diagonal and offdiagonal blocks $M_d$ and $M_o$. In this basis, the two matrices become diagonal, where the diagonal elements are given by the eigenvalues
\beeq
\label{eq:dwmateig}
(\l_\txt{d},\l_\txt{o})=(g_2,g_3),(g_2-N_fg_1,-(N_f-1)g_3).
\eneq
These two sets of eigenvalues are $N_f-1$ fold and singly degenerate, respectively.

The $N_f-1$ fold degenerate eigenstates correspond to the flavor density waves (FDW) with symmetric flavor pairings. Each channel exhibits the corrections
\beeq
\label{eq:rgeqfdw}
\fr{d}{dy}\lf(\bear{c}\D_\txt{FDW}\\\D_\txt{FDW}^*\enar\ri)=d_1\lf(\bear{cc}g_2&g_3\\g_3&g_2\enar\ri)\lf(\bear{c}\D_\txt{FDW}\\\D_\txt{FDW}^*\enar\ri),
\eneq
where the diagonal and offdiagonal elements result from the eigenvalues Eq.~(\ref{eq:dwmateig}) of original blocks $M_d$ and $M_o$. The eigenstates represent the real and imaginary FDW channels with symmetric flavor pairings. Each eigenvalue is identified with the interaction
\beeq
\label{eq:testvertexpfdw}
g_\txt{r/iFDW}=-(g_2\pm g_3).
\eneq
Note that we have introduced an extra minus sign for consistency with the definition of interaction in Eq.~(\ref{eq:tvcorrection}). On the other hand, the singly degenerate eigenstate of $M$ is recognized as the charge density wave (CDW). The correction matrix for CDW is determined by the singly degenerate eigenvalues in Eq.~(\ref{eq:dwmateig})
\beeq
\beal
\fr{d}{dy}\lf(\bear{c}\D_\txt{CDW}\\\D_\txt{CDW}^*\enar\ri)&=d_1\lf(\bear{cc}g_2-N_fg_1&-(N_f-1)g_3\\-(N_f-1)g_3&g_2-N_fg_1\enar\ri)
\\
&\quad\times\lf(\bear{c}\D_\txt{CDW}\\\D_\txt{CDW}^*\enar\ri).
\enal
\eneq
Diagonalizing this equation, we arrive at the interactions in the real and imaginary CDW channels
\beeq
g_\txt{r/iCDW}=N_fg_1-g_2\pm(N_f-1)g_3.
\eneq
Again we have introduced an extra minus sign for consistency with the definition of interaction in Eq.~(\ref{eq:tvcorrection}).

The density wave channels with pairing between different flavors $\s\neq\s'$ are the FDW channels with antisymmetric flavor pairings. With the test vertices
\beeq
\d H=\sum_{\s\neq\s'}(\D_\txt{FDW}\psi_{\b\s}^\dag\psi_{\a\s'}+\D_\txt{FDW}^*\psi_{\a\s'}^\dag\psi_{\b\s})
\eneq
where $\a>\b$, we find the same correction equation as Eq.~(\ref{eq:rgeqfdw}). The same interaction Eq.~(\ref{eq:testvertexpfdw}) as the FDW channels with symmetric flavor pairings is obtained.

\section{Landau-Ginzburg theory for loop current order}
\label{app:lg}

The dominance of loop current order is observed near Van Hove filling. Here we derive the dominant loop current order through a Landau-Ginzburg analysis, where the minimization of free energy determines the leading state below the critical temperature $T_c$.

The analysis starts by projecting the low energy theory onto CDW channel. In this channel, the interactions $g_1$, $g_2$, $g_3$ contribute, while $g_4$ does not. With the Fierz identity, the quartic interactions are projected onto the channel of CDW bilinears $\psi_\a^\dag\psi_\b$ with $\a>\b$
\beeq
\beal
H_\txt{int}
&=\fr{1}{4N_f}\sum_{\a>\b}\bigg\{2(N_fg_1-g_2)(\psi_\a^\dag\psi_\b)(\psi_\b^\dag\psi_\a)\\
&~+(N_f-1)g_3\lf[(\psi_\a^\dag\psi_\b)(\psi_\a^\dag\psi_\b)+(\psi_\b^\dag\psi_\a)(\psi_\b^\dag\psi_\a)\ri]\bigg\}.
\enal
\eneq
Decompose the interactions further into real and imaginary parts, with the bilinears defined as $(\psi_\a^\dag\psi_\b)_{r/i}=(\psi_\a^\dag\psi_\b\pm\psi_\b^\dag\psi_\a)/2$. The loop current channel is identified as the imaginary channel
\beeq
H_\txt{int}
=\fr{g_\txt{iCDW}}{2N_f}\sum_{\a>\b}(\psi_\a^\dag\psi_\b)_i^\dag(\psi_\a^\dag\psi_\b)_i,
\eneq
where the interaction $g_\txt{iCDW}$ diverges in the negative direction under the RG flow.

A Hubbard-Stratonovich transformation is performed so that the quartic interaction is decomposed by the bosonic auxiliary fields. The action takes the form
\beeq
\beal
S
&=\int_0^{1/T} d\tau\Bigg\{\sum_\a\psi_\a^\dag(\p_\tau+\xi_\a)\psi_\a+\fr{2N_f}{|g_\txt{iCDW}|}\sum_{\a>\b}\til\D_{\a\b}^2\\
&\quad-\sum_{\a>\b}i\til\D_{\a\b}\lf[(\psi_\a^\dag\psi_\b)_i^\dag-(\psi_\a^\dag\psi_\b)_i\ri]\Bigg\},
\enal
\eneq
where the order parameter is defined as $\D_{\a\b}=i\til\D_{\a\b}$ with real $\til\D_{\a\b}$. Impose the static condition $\til\D_{\a\b}(\tau)=\til\D_{\a\b}$. After a temporal Fourier transform $\psi(\tau)=\sqrt T\sum_n\psi_ne^{-i\o_n\tau}$, the mean field free energy reads
\beeq
\beal
F
&=\fr{2N_f}{|g_\txt{iCDW}|}\sum_{\a>\b}\til\D_{\a\b}^2-T\sum_n\sum_\a\psi_\a^\dag G_\a^{-1}\psi_\a\\
&\quad+T\sum_n\sum_{\a>\b}i\til\D_{\a\b}\lf(\psi_\a^\dag\psi_\b-\psi_\b^\dag\psi_\a\ri).
\enal\eneq
Consider the hexagonal lattices $N_p=3$. Express the free energy in a matrix representation of patch basis
\beeq
F=\fr{2N_f}{|g_\txt{iCDW}|}|\til{\mbf\D}|^2-T\sum_n\intv{k}\psi^\dag\mca G^{-1}\psi,
\eneq
where the inverse loop current propagator is defined
\beeq
-\mca G^{-1}=\lf(\bear{ccc}-G_1^{-1}&-i\til\D_3&i\til\D_2\\i\til\D_3&-G_2^{-1}&-i\til\D_1\\-i\til\D_2&i\til\D_1&-G_3^{-1}\enar\ri).
\eneq
Integrating out the fermion field, we obtain the free energy as a function of order parameter
\beeq
F=\fr{2N_f}{|g_\txt{iCDW}|}|\til{\mbf\D}|^2-\Tr\ln(-\mca G^{-1}),
\eneq
where $\ln\det(-\mca G^{-1})=\Tr\ln(-\mca G^{-1})$ is utilized.

When the temperature is just below the critical temperature $T\lesssim T_c$, the order parameter $\til{\mbf\D}$ is infinitesimal. A Landau-Ginzburg analysis can be conducted in this regime. Define $\mca G_0=\mca G(\til{\mbf\D}=0)$ and $\hat\D=(-\mca G^{-1})-(-\mca G_0^{-1})$. With the constant part ignored, an expansion up to quartic order of $\til{\mbf\D}$ is obtained
\beeq
F=\fr{2N_f}{|g_\txt{iCDW}|}|\til{\mbf\D}|^2+\fr{1}{2}\Tr(-\mca G_0\hat\D)^2+\fr{1}{4}\Tr(-\mca G_0\hat\D)^4.
\eneq
The quadratic order takes the form $\a(T-T_c)|\til{\mbf\D}|^2$ with $\a>0$. As $T<T_c$, the coefficient becomes negative, and the symmetry broken states arise. For the quartic order
\beeq
\fr{N_f}{2}\bigg[Z_1|\til{\mbf\D}|^4+2(Z_2-Z_1)\lf(\til\D_1^2\til\D_2^2+\til\D_2^2\til\D_3^2+\til\D_3^2\til\D_1^2\ri)\bigg],
\eneq
the coefficients manifest the `square diagrams' \cite{nandkishore12prl}
\beeq
Z_1=\Tr(G_1^2G_2^2),\quad Z_2=\Tr(G_1^2G_2G_3).
\eneq
The calculations indicate $Z_1>0$, which guarantees the stability of Landau-Ginzburg theory at quartic order. Meanwhile, the second diagram is $0<Z_2<Z_1$. Since $Z_2-Z_1<0$, the loop current order develops at all three nesting momenta simultaneously, known as a $3Q$ state.

\bibliography{Reference}

\end{document}